\documentclass[twocolumn]{aastex63}
\usepackage{graphicx}
\usepackage{amsmath}
\usepackage{appendix}
\usepackage{rotating}
\begin{document}

\title{Joint Radial Velocity and Direct Imaging Planet Yield Calculations: \\ I. Self-consistent Planet Populations} 

\author{Shannon D. Dulz}
\affiliation{Department of Physics, University of Notre Dame, 225 Nieuwland Science Hall, Notre Dame, IN, 46556, USA}

\author{Peter Plavchan} 
\affiliation{Department of Physics and Astronomy, George Mason University, 4400 University Drive MS 3F3, Fairfax, VA, 22030, USA}

\author{Justin R. Crepp}
\affiliation{Department of Physics, University of Notre Dame, 225 Nieuwland Science Hall, Notre Dame, IN, 46556, USA}

\author{Christopher Stark}
\affiliation{Space Telescope Science Institute, 3700 San Martin Drive, Baltimore, MD 21218}

\author{Rhonda Morgan}
\affiliation{Jet Propulsion Laboratory, California Institute of Technology, 4800 Oak Grove Drive, Pasadena, CA 91109}

\author{Stephen R. Kane}
\affiliation{Department of Earth and Planetary Sciences, University of California, 900 University Avenue, Riverside, CA 92521}

\author{Patrick Newman}
\affiliation{Department of Physics and Astronomy, George Mason University, 4400 University Drive MS 3F3, Fairfax, VA, 22030, USA}

\author{William Matzko}
\affiliation{Department of Physics and Astronomy, George Mason University, 4400 University Drive MS 3F3, Fairfax, VA, 22030, USA}

\author{Gijs D. Mulders}
\affiliation{Department of the Geophysical Sciences, University of Chicago, 5734 S. Ellis Avenue, Chicago, Illinois 60637}
\affiliation{Earths in Other Solar Systems Team, NASA Nexus for Exoplanet System Science}

\correspondingauthor{Shannon D. Dulz}
\email{sdulz@nd.edu}

\begin{abstract}
Planet yield calculations may be used to inform the target selection strategy and science operations of space observatories. Forthcoming and proposed NASA missions, such as the Wide-Field Infrared Survey Telescope (WFIRST), the Habitable Exoplanet Imaging Mission (HabEx), and the Large UV/Optical/IR Surveyor (LUVOIR), are expected to be equipped with sensitive coronagraphs and/or starshades. We are developing a suite of numerical simulations to quantify the extent to which ground-based radial velocity (RV) surveys could boost the detection efficiency of direct imaging missions. In this paper, we discuss the first step in the process of estimating planet yields: generating synthetic planetary systems consistent with observed occurrence rates from multiple detection methods. In an attempt to self-consistently populate stars with orbiting planets, it is found that naive extrapolation of occurrence rates (mass, semi-major axis) results in an unrealistically large number-density of Neptune-mass planets beyond the ice-line ($a \gtrsim 5$au), causing dynamic interactions that would destabilize orbits. We impose a stability criterion for multi-planet systems based on mutual Hill radii separation. Considering the influence of compact configurations containing Jovian-mass and Neptune-mass planets results in a marked suppression in the number of terrestrial planets that can exist at large radii. This result has a pronounced impact on planet yield calculations particularly in regions accessible to high-contrast imaging and microlensing. The dynamically compact configurations and occurrence rates that we develop may be incorporated as input into joint RV and direct imaging yield calculations to place meaningful limits on the number of detectable planets with future missions.
\end{abstract}

\section{Introduction}\label{sec:intro}

The sensitivity and multiplexing capabilities of planet-finding telescopes and instruments continues to improve. As different detection methods systematically explore and discover new worlds, the parameter space within which planets reside unfolds to reveal a complex and rich structure with properties and correlations previously hidden from view. Statistical distributions provide insights into the physics of planet formation, evolution, and relation to host star properties \citep{Winn2015}. If a broad understanding of when, how, and where planets tend to take up residence could be obtained, it might point to effective methods for detecting worlds that are most likely to develop and sustain life. The atmospheres of such worlds could then be studied in exquisite detail and searched for potential biosignatures using stable, direct imaging platforms from space \citep{Seager2015,Stapelfeldt2014}. 

Space-based imaging missions will rely upon exoplanet statistics available in the years preceding launch to help develop an optimal observing strategy \citep{NAS2018}. Demographics information will be informed from a variety of large-scale surveys, many of which were conducted using observations of nearby stellar systems \citep{Johnson2010,Bowler2015,Stone2018,Nielsen2019}. Although each planet-detection technique confers different observational biases, complementary sensitivities such as radial velocity (RV) and high-contrast imaging also present an opportunity to combine results in an effort to form a more holistic picture of exoplanetary system architectures \citep{Montet2014,Kane2019}. 

This paper represents the first in a series of articles that attempt to quantify the extent to which ground-based RV observations might facilitate the scientific yield and efficiency of forthcoming direct-imaging space missions. To simulate surveys, we must utilize our current best quantitative knowledge of occurrence rates for planets that such surveys will attempt to detect in the first-place, presenting an iterative ``Catch-22" of sorts. Ideally, we would know the occurrence of planets as a function of relevant parameters, such as mass and semi-major axis, perfectly so that we can optimally design a survey to detect and characterize a scientifically compelling sample size. However, uncertainties in planet occurrence and reliance on extrapolation obfuscate the survey strategy and required scale.

Obtaining a comprehensive, yet unbiased, empirical view of exoplanet ``demographics" is currently limited by the inherent sensitivity of detection methods, available telescopes and instruments, observational biases, and selection effects. For instance, the transit method is sensitive to planets large (Jovian) and small (terrestrial), but primarily at short orbital periods \citep[e.g. $<\:\sim$1 yr,][]{Kopparapu2018}. Direct imaging, radial velocity, microlensing, and astrometric methods are sensitive to longer-period planets relative to \emph{Kepler}, but are currently limited to more massive planets; direct imaging is sensitive to young Jovians orbiting A-stars, and microlensing to planets near the Einstein ring radius of  M-stars \citep[e.g.,][]{Crepp2011,Perryman2014ApJ,Bryan2016,Cumming2008,Fernandes2019,Santerne2018,Clanton2016}.

Beyond the time baseline of the \emph{Kepler} mission and beyond the mass sensitivity limits of RV, direct imaging, astrometry and microlensing surveys, empirical demographics information at large ($>$1 au) orbital distances for low-mass ($< 1 M_J$) planets becomes largely incomplete for F, G, and K-type stars \citep[e.g.,][]{Fernandes2019,Mulders2018, Howard2012}. Exploring small planet populations at large semi-major axes is thus currently reliant upon extrapolation, theory, and simulation \citep[e.g., ][]{Mordasini2018}. As shown by \citet{Burke2015}, the method of extrapolation can significantly impact estimated occurrence rates. Extrapolation also does not take into account the physical dynamic limitations of systems. 

To provide a comprehensive picture of the state of demographics knowledge gained from \emph{Kepler}, the NASA Exoplanet Exploration Program Analysis Group (ExoPAG) Study Analysis Group (SAG) 13 recently compiled over one dozen, community-sourced, transiting planet occurrence rate tables from both peer-reviewed studies \citep{Burke2015,DressingCharbonneau2013,Foreman-Mackey2014,Petigura2013,Traub2015} and also unpublished \emph{Kepler} ``hack week'' analyses \citep{SAG13,Kopparapu2018}. These sources were then reprocessed over a standardized radius-period grid \citep{Kopparapu2018}. The meta-analysis completed by SAG 13 represents a current field-wide consensus on occurrence rates for short period planets (10-640 days) across a broad range in planetary radii.  More recent and more precise exoplanet radii measurements are available by making use of improved stellar radii determinations from \textit{Gaia} \citep{Fulton2018,Berger2018}, to update \textit{Kepler} planet occurrence rate calculations \citep{Pascucci2019,Shabram2019}.
There has, however, not yet been an updated and revised comprehensive analysis of community-sourced exoplanet demographics analogous to SAG13.  As such, we use SAG 13 rates as the first step in simulating realistic planetary systems for mission yield simulations.  

In this paper, we compare the results from extrapolating planet population models based on SAG 13 in combination with several radial velocity based models with additional approximate dynamic stability criteria based on mutual Hill radii. We identify inconsistencies between the extrapolated occurrence of planets and the number that can remain stable. In the near future, this phase space will continue to be probed with a variety of surveys from the ground and space \citep{NAS2018}. In this paper, we use simple stability criteria to draw reasonable, useful upper limits on the occurrence rates of small planets at long periods.

The article is organized as follows: In $\S$~\ref{sec:methods}, we summarize previous demographics results that form the basis of our analysis as well as detail the parameter space over which they are combined. We then describe an approach for drawing planets from demographics with a summary of our calculations of the total number of planets per star. We detail a procedure for drawing randomized planet mass and semi-major axis parameters from joint distributions. We further introduce dynamical stability criteria we use to generate realistic systems. In $\S$~\ref{sec:results} we parameterize the effect of these stability criteria and their use in extending planet demographics beyond their original validity. Lastly, $\S$~\ref{sec:conclusions} summarizes the results and their implications for imaging follow-up missions. 

\section{Methods}
\label{sec:methods}

\subsection{Demographic Sets}\label{sec:sets}

We construct ``demographic sets" in mass and semi-major axis space  ($M_p, a$) from recent demographic studies in the literature. Multiple data sets are used to cover a broad span of planet parameters. These studies focused on F, G, and K-type stars, and consequently we limit the scope of this analysis to F, G, and K-type stars as well and assume 1.0 $M_{\rm \odot}$ stars. We define three comparative distributions, referred to as SAG13, RV1, and RV2. While all three demographics sets utilize SAG 13 occurrence rates for Neptune and Terran planets, differing the treatment of Jovians in each set allows us to investigate the effect these largest planets have on the stability of the entire demographics parameter space. The sets are summarized as:

\begin{itemize}
    \item SAG13 utilizes a two-part power-law based on occurrence rates derived from the \emph{Kepler} space mission, using Eqn 4 from \citet[][]{Kopparapu2018};
    \item RV1 replaces SAG13 Jovian occurrence rates in the region $0.225 \leq m_p / M_J \leq 15$ with a combination of \citet{Bryan2016} and \citet{Cumming2008}, while retaining SAG13 demographics for the least massive planets; 
    \item RV2 replaces SAG13 Jovian occurrence rates in the region $0.225 \leq m_p / M_J \leq 15$ with \citet[][]{Fernandes2019}, while retaining SAG13 demographics for the least massive planets.
\end{itemize}

The rates derived by \citet[][]{Cumming2008} are based on 8 years of Keck Planet Search radial velocity data for F, G, and K-type stars. \citet[][]{Cumming2008} rates were originally constrained to $m_p\geq 0.3M_J$ (which we extrapolate slightly to $0.225M_J$) and to periods $\leq$ 2000 days $\approx$ 3.1 au (which we extrapolate to 5 au).

In the RV1 set, complementary to the rates for close-in Jovians from \citet{Cumming2008}, we use \citet{Bryan2016} rates for cold Jupiters. \citet{Bryan2016} utilized a Keck Doppler survey in combination with NIRC2 AO imaging to derive an occurrence rate of cold Jovian companions to systems with known radial velocity detected planets (primarily hot or warm Jovian planets). \citet{Bryan2016} calculated occurrence rates based on fits to several mass and semi-major axes ranges from which we have chosen to use the rates valid for $ 0.5-13 M_J$ (which we extend down to $0.225M_J$) and $5-50$ au. \citet{Bryan2016} derived their demographics for Jovian planets at large semi-major axes from hot-Jupiter hosting stars. For the purposes of this study, we assume these demographics for cold Jupiters apply for all stars, regardless of whether or not they host a hot or warm Jupiter. In other words, we assume the hot/warm and cold Jupiter populations are decoupled.  One could expect that these planet populations are instead correlated, and that stars with hot Jupiters are more likely to host cold Jupiters as well. It is not yet known definitively whether or not these two planet populations are coupled or not.

For the RV2 set, we utilize the \citet{Fernandes2019} meta-analysis of Jovian occurrence rates. \citet{Fernandes2019} re-derived occurrence rates based on the results and sensitivity of Kepler-based surveys \citep{Mathur2017,Thompson2018} with the HARPS and CORALIE RV survey from \citet{Mayor2011} and identified a power-law break in the occurrence rates at $P \approx 2075$ days. \citet{Fernandes2019} present a parametric distribution of occurrence rates for planets $0.1-20 M_J$ that we extend to the $0.06-35$ au range.
Differences in the Jovian planet population could play an essential role in shaping the overall dynamic architecture of planetary systems. The latter sets, RV1 and RV2, are based on archival Doppler RV measurements with time baselines of several decades. By exploring (non-transiting) giant planets at larger semi-major axes, RV1 and RV2 both probe a planet mass space that is degenerate in planet radii in SAG13, and also probe a time baseline longer than the Kepler survey duration. 

For simplicity, planet occurrence rates have generally been described using power-laws in the literature (e.g. \citealt{Johnson2010,Cumming2008}). Given a sufficient number of detections, more complex structure can be observed, both at short orbital periods of $\lesssim$10 days \citep[]{Youdin2011}, and several studies now report evidence for breaks in the best-fitting power-law distributions at semi-major axes of a few au \citep{Suzuki2016,Fernandes2019,Meyer2018}. Indeed, a number of power-law breaks occur in the planet distributions we consider which, if real, could provide insights into planet formation mechanisms. For the purposes of this study, deviations from a single power-law are also relevant because they impact system dynamic stability when extrapolating to large semi-major axes (see below).  This in turn impacts the likelihood for direct detection for future space-based missions and mission concepts such as WFIRST, HabEx and LUVOIR \citep[][]{WFIRSTfinal,HabExfinal,LUVOIRfinal}. 

\begin{figure}[t]
\gridline{\fig{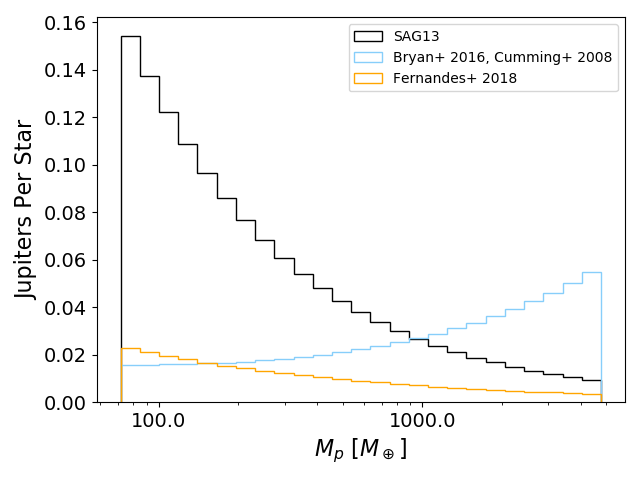}{0.48\textwidth}{}}
\gridline{\fig{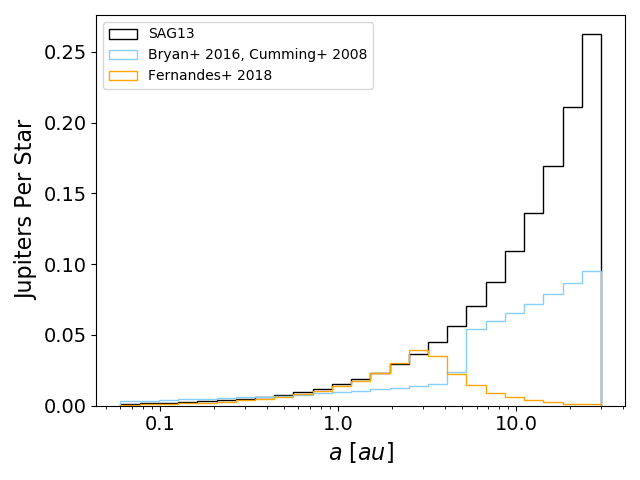}{0.48\textwidth}{}}
\caption{Occurrence rate per star predicted by the \citet[][]{SAG13} ``nominal," \citet[][]{Bryan2016} and \citet[][]{Cumming2008}, and \citet[][]{Fernandes2019} data sets. The average number of Jovian planets in the range $0.225 \leq m / M_
J \leq 15$ is plotted as a function of planet mass (top panel) and semi-major axis (bottom panel). The transition from \citet[][]{Cumming2008} to \citet[][]{Bryan2016} at $a \approx 5$ au is visible as an abrupt increase in the distribution due to selection effects. Contrasting this result, at $a \approx 3$ au \citet[][]{Fernandes2019} find a turn-over in the Jovian population.}
\label{fig:JupiterTheory1D}
\end{figure}

Figure~\ref{fig:JupiterTheory1D} displays the planet mass and semi-major axis dependence of planet occurrence rates for (only) gas giant planets after converting to common units (see ${\S}$\ref{sec:combining}). SAG13 and RV2 predict a negative, power-law relationship with planet mass, whereas RV1 predicts an increasing occurrence rate with increasing planet mass. In semi-major axis (or orbital period), SAG13 and RV1 predict increasing occurrence rates for cold Jupiters, with a noticeable break at 5 au where we transition from \citet[][]{Cumming2008} to \citet[][]{Bryan2016}). RV2, however, predicts a turn-over around $a \approx 3$ au corresponding to $P \approx 2075$ days for solar-type stars.

\begin{figure*}[t]
\gridline{\fig{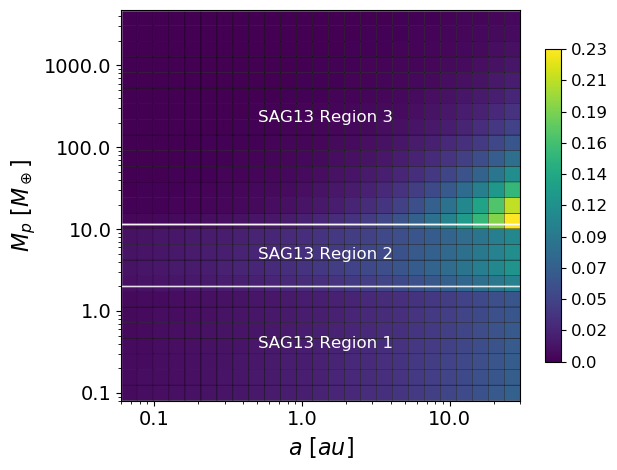}{0.33\textwidth}{SAG 13}
          \fig{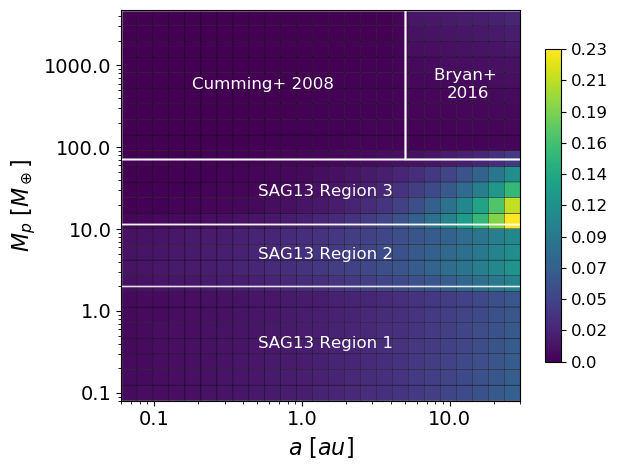}{0.33\textwidth}{RV 1}
          \fig{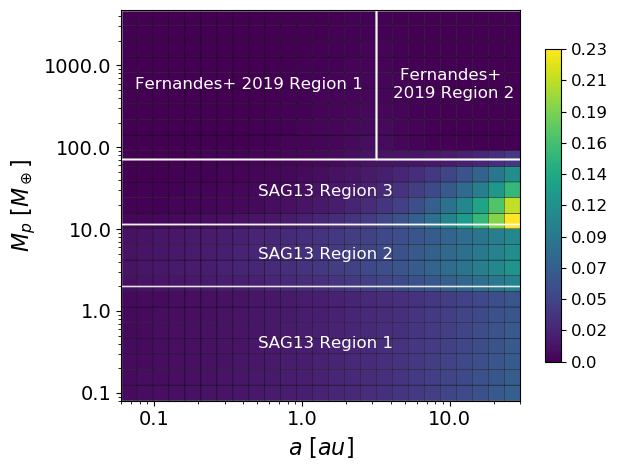}{0.33\textwidth}{RV 2}}
\caption{Visual comparison of demographic sets. Planet occurrence is quantified using the power-law distributions and integrals described in the text. The results are shown in units of planets per star per two-dimensional (mass-semi-major axis) bin. SAG13 identified two different regimes in period-radius space; however, when converting to mass with the Chen \& Kipping mass-radius relation, an additional break in mass is necessarily introduced  (see ${\S}$2.2.1).}
\label{fig:Theory2D}
\end{figure*}
      
\subsection{Combining Demographic Sets}\label{sec:combining}
Given a compilation from multiple literature references and composite data sets, we attempt to self-consistently combine their results to fill in the mass-semi-major axis plane. We take into account the regions over which different demographics are sensitive, extrapolating only in the case of SAG13 and large semi-major axes where empirical demographics knowledge is least constrained. 

We employ analytic distributions parameterized in the form of a double power-law. While units and notation vary between planet detection methods and different publications, they can generally be quantified in the form:
\begin{equation}
\frac{d^2 N}{d \log_Z X \; d \log_Z Y} = C_{X,Y,Z} X^{\alpha} \: Y^{\beta},
\end{equation} 
\noindent where $X$ represents either planet mass or radius, $Y$ represents planet semi-major axis or orbital period, and $Z$ denotes the logarithm in either base-$e$ or base-10. In what follows, masses are given in either Earth masses or Jupiter masses; radii are given in Earth radii; semi-major axes are given in au; and period is given in either days or years. The coefficient $C_{X,Y,Z}$ governing the number density of planets has corresponding units.

In any particular variation of the demographics, each power-law is applied over a given mass and semi-major axis range, notated as $M_{p_{\rm min}}=0.08M_{\rm\oplus}$ to $M_{p_{\rm max}}=15M_{\rm Jup}$ and $a_{\rm min}=0.06$ au to $a_{\rm max}=35$ au. These region limits can be directly translated into the units of each power-law variation, $X_{\rm min}$ to $X_{\rm max}$ and $Y_{\rm min}$ to $Y_{\rm max}$. We choose this mass lower limit to allow for direct comparison to the findings of \citet{Kopparapu2018} which we explore in $\S$~\ref{sec:results}.  We choose this semi-major axis minimum to avoid the demographics break interior to this location that implies additional physics due to tidal dissipation and stellar irradiation for the closest known planets \citep[e.g.][]{Youdin2011,Petigura2013,Howard2012}.  We set the mass upper-limit of 15 $M_{\rm Jup}$ to the approximate Jovian--brown dwarf boundary, and we somewhat arbitrarily set the semi-major axis upper-limit to 35 au, just beyond Neptune in in our Solar System.

We first determine the number of planets expected in a region by integrating within the mass and semi-major axis limits: 
\begin{multline}
    N_{\rm region} = \\
    \int \limits_{\log_Z (X_{\rm min})}^{\log_Z (X_{\rm max})} \int\limits_{\log_Z (Y_{ \rm min})}^{\log_Z (Y_{\rm max})} C_{X,Y,Z} X^{\alpha_X}  \: Y^{\beta_Y} d \log_Z X \; d \log_Z Y.
\end{multline}

Evaluating the integral yields:
\begin{equation}
    N_{\rm region} = \frac{C_{X,Y,Z}}{\alpha_X \beta_Y (\ln{Z})^2}(X_{\rm max}^{\alpha_X} - X_{\rm min}^{\alpha_X})(Y_{\rm max}^{\beta_Y} - Y_{\rm min}^{\beta_Y}).
    \label{eq:intergratedgeneral}
\end{equation}

Figure~\ref{fig:Theory2D} shows a two-dimensional view of the demographics sets used as input to this study using this formalism. Several regions of interest in mass and semi-major axis space over which power-law breaks occur, and/or the distributions transition from one study to another, are delineated using white lines. We naively and intentionally extrapolate power-law distributions to the edges of these regional boundaries (see $\S$\ref{sec:results} for further discussion).  

\subsubsection{Mass-Radius Relation for SAG13}
We use the mean mass-radius relation from \citet{ChenKipping2017} to convert planet occurrence rates measured in units of planet radius to units of planet mass. 
The SAG13 distribution is a two-part power-law with a break at $3.4 \,R_{\oplus}$, corresponding to an average radius of $11.5\, M_{\oplus}$. The break defines the division between SAG13 Region 2 and Region 3 as we have defined them in Fig.~\ref{fig:Theory2D}. We further divide the SAG13 set into Region 1 and Region 2 at $2.04\, M_{\oplus}$ (Fig.~\ref{fig:Theory2D}), corresponding to the different mass-radius distributions for ``Terran" and ``Neptunian" worlds in \citet{ChenKipping2017}.

\subsubsection{Semi-major axis boundaries for RV1 and RV2}
In data set RV1, the transition from \citet{Cumming2008} to \citet{Bryan2016} occurs at $a=5$ au, which is slightly extrapolated from the 2000-day ($\sim$3.1 au for solar mass stars) maximum period explored in \citet[]{Cumming2008}, and the minimum semi-major axis explored in \citet{Bryan2016}. Combining the \citet{Cumming2008} and \citet{Bryan2016} data sets results in a distinct jump (offset) in Jovian planet occurrence, likely due to the \citealt{Bryan2016} sample including only stars with known short-period RV planets (Fig.~\ref{fig:JupiterTheory1D}). We do not attempt to correct/scale/align the jump at this boundary for simplicity, effectively assuming these hot and cold Jovian planet populations are uncorrelated.  In data set RV2, \citet{Fernandes2019} instead finds an occurrence rate turnover at $P = 2045$ days ($a \approx 3.15 $ au for solar mass stars).We denote either side of the power-law break in semi-major axis from \citet{Fernandes2019} as Region 1 and Region 2 (Fig.~\ref{fig:Theory2D}).  

\subsubsection{Parameterizing Uncertainties}
We do not consider the 1-$\sigma$ uncertainties in the power law exponents and proportionality constants, which quickly becomes multi-dimensional for the regions comprising a demographics set. Rather, as done in \citet{SAG13}, given the dominant remaining uncertainty in planet occurrence as a function of planet mass and orbital radius, it is instructive to consider different ``optimistic'' (lots of planets) and ``pessimistic" (fewer planets) outcomes, relative to ``nominal" distributions. These various possible outcomes are listed in Table~\ref{tab:demosets} to represent the compiled power-laws that we consider. As done with the SAG13 analysis, these categories are useful for establishing approximate upper and lower confidence intervals to the number of planets in a given parameter space. For the RV1 and RV2 sets, we do not consider variations in the demographic uncertainties for Jovian planets as they are formulated alternatively as uncertainties on the power law exponents and constants.

\begin{deluxetable*}{l c c c c c c c c c c}\label{tab:demosets}
    \tablecaption{Demographics sets used for simulations.}
    \tablehead{\colhead{Reference}  & \colhead{$\alpha_X$} & \colhead{$\beta_Y$} & \colhead{$C_{X,Y,Z}$} & \colhead{$X$} & \colhead{$Y$} & \colhead{$Z$} & \colhead{$M_{p_{\rm min}}$} & \colhead{$M_{p_{\rm max}}$} & \colhead{$a_{\rm min}$} & \colhead{$a_{\rm max}$}}
    \startdata
    Bryan et al. (2016) & 0.56 & 0.38 & 0.015  & $M_{\rm Jup}$ & $a_{\rm au}$ & $e$ & 75.51\tablenotemark{c} & 4767.43\tablenotemark{d} & $5$ & $35$\\
    Cumming et al. (2008) & -0.31 & 0.26 & 0.00146 & $M_{\rm Jup}$ & $P_{\rm days}$ & $e$ & 75.51\tablenotemark{c} & 4767.43\tablenotemark{d} & $0.06$ & $5$\\
    Fernandes et al. (2019), Region 1 & -0.46 & 0.7 & 0.0114 \tablenotemark{a}& $M_{\oplus}$ & $P_{\rm days}$ & $e$ & 75.51\tablenotemark{c} & 4767.43\tablenotemark{d} & $0.06$ & 3.18\tablenotemark{e}\\
    Fernandes et al. (2019), Region 2 & -0.46 & -1.2 & 22882.3 \tablenotemark{b} & $M_{\oplus}$ & $P_{\rm days}$ & $e$ & 75.51\tablenotemark{c} & 4767.43\tablenotemark{d} & 3.18\tablenotemark{e} & $35$\\
    SAG13, Region 1, pessimistic & 0.277 & 0.204 & 0.138  & $R_{\oplus}$ & $P_{\rm yr}$ & $e$ & $0.08$ & $2.04$ & $0.06$ & $35$\\
    SAG13, Region 2, pessimistic & 0.277 & 0.204 & 0.138 & $R_{\oplus}$ & $P_{\rm yr}$ & $e$ & $2.04$ & $11.467$ & $0.06$ & $35$  \\
    SAG13, Region 3, pessimistic & -1.56 & 0.51  & 0.72 & $R_{\oplus}$ & $P_{\rm yr}$ & $e$ & $11.467$ & 75.51\tablenotemark{c} or 4767.43\tablenotemark{d} & $0.06$ & $35$\\
    SAG13, Region 1, nominal     & -0.19 & 0.26  & 0.38 & $R_{\oplus}$ & $P_{\rm yr}$ & $e$ & $0.08$ & $2.04$ & $0.06$ & $35$\\
    SAG13, Region 2, nominal     & -0.19 & 0.26  & 0.38 & $R_{\oplus}$ & $P_{\rm yr}$ & $e$ & $2.04$ & $11.467$ & $0.06$ & $35$  \\
    SAG13, Region 3, nominal     & -1.18 & 0.59  & 0.73 & $R_{\oplus}$ & $P_{\rm yr}$ & $e$  & $11.467$ & 75.51\tablenotemark{c} or 4767.43\tablenotemark{d} & $0.06$ & $35$\\
    SAG13, Region 1, optimistic  & -0.68 & 0.32  & 1.06 & $R_{\oplus}$ & $P_{\rm yr}$ & $e$ & $0.08$ & $2.04$ & $0.06$ & $35$\\
    SAG13, Region 2, optimistic  & -0.68 & 0.32  & 1.06 & $R_{\oplus}$ & $P_{\rm yr}$ & $e$ & $2.04$ & $11.467$ & $0.06$ & $35$  \\
    SAG13, Region 3, optimistic  & -0.82 & 0.67  & 0.78 & $R_{\oplus}$ & $P_{\rm yr}$ & $e$  & $11.467$ & 75.51\tablenotemark{c} or 4767.43\tablenotemark{d} & $0.06$ & $35$\\
   \enddata
   \tablenotetext{a}{$C_{X,Y,Z}=\frac{0.83}{10^{-0.46}2075^{0.7}}$}
   \tablenotetext{b}{$C_{X,Y,Z}=\frac{0.83}{10^{-0.46}2075^{-1.2}}$}
   \tablenotetext{c}{$0.225M_J$}
   \tablenotetext{d}{$15M_J$} 
    \tablenotetext{e}{Period $P=2075$ days converted to au assuming a $M=1.0M_{\odot}$ star.}
\end{deluxetable*}

\subsection{Drawing Planets to Create Populations}\label{sec:drawing}

Synthetic planetary systems are generated based on the chosen demographic models. To remain consistent with the catalogues used to derive the SAG13, RV1, and RV2 distributions, we consider single stars comparable in mass to the Sun. A total of 100,000 planetary systems are generated, and the results statistically averaged. Stars are populated with planets by randomly drawing from these demographic models. The resulting orbital configurations are then checked for stability using semi-analytic methods.

\begin{figure}
    \centering
    \includegraphics[width=0.5\textwidth]{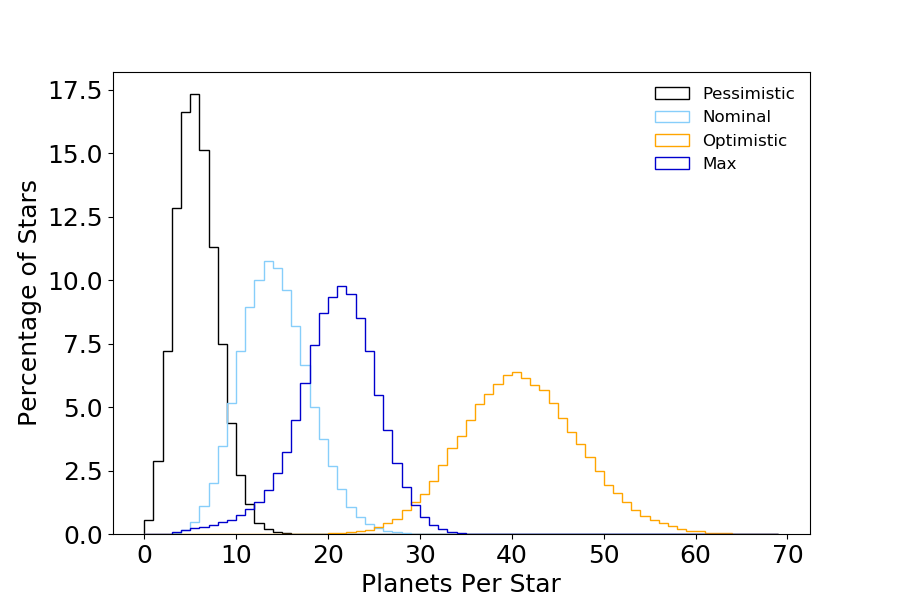}
    \caption{Planets per star over the entire parameter space for the optimistic, nominal, and pessimistic populations. The optimistic scenario includes an inordinate number of Neptune-mass and lower-mass planets. We show that the ``optimistic" scenario taken from SAG13 is dynamically unstable by comparison to the maximally dynamically packed scenario discussed in $\S$\ref{sssec:maxpack}.}
    \label{fig:PlanetsPerStar}
\end{figure}

The expected number of planets for a given region of mass and semimajor axis space of a given star is drawn from an assumed Poisson distribution \citep{Ballard2016,Lissauer2014}, 
\begin{equation}\label{eq:poisson}
    P(k) = \frac{N_{\rm region}^k e^{-N_{\rm region}}}{k!},
\end{equation}
where $P(k)$ is the probability that a star has $k$ integer planets per region. We do not apply an additional multiplicative factor representing the fraction of stars with planets -- e.g. that there exists a population of stars with no planets in excess of a Poisson distribution, as is sometimes done in the literature \citep[]{Mulders2018,Zhu2018,He2019}. In other words, we effectively assume the fraction of stars with planets per region is unity, and that the average number of stars per planet per region holds for all planets and that the individual number of planets per star per region follows a Poisson distribution which can include zero planets (see also \ref{sec:conclusions}). The value of $N_{\rm region}$ is found by evaluating Equation~\ref{eq:intergratedgeneral} when integrating over mass and semi-major axis for each distinct region of the chosen demographics set,
\begin{equation}
    N_{\rm region} = \frac{C}{\alpha\beta}(M_{p_{\rm max}}^{\alpha} - M_{p_{\rm min}}^{\alpha})(a_{\rm max}^{\beta} - a_{\rm min}^{\beta}).
    \label{eq:Npl}
\end{equation}
For example, Region 2 of SAG13 has boundaries $ 0.06\leq a / au \leq 35 $ and $2.04 \leq M_p / M_{\rm \oplus} \leq 11.467$ (Tab.~\ref{tab:demosets}). Figure~\ref{fig:PlanetsPerStar} shows an example histogram of the SAG13 distributions.

For each individual planet, we then draw a mass and semi-major axis from the joint probability distribution using ``inverse transform sampling" in each direction, which enables random sampling from a cumulative distribution. Upon generating two uniformly random numbers, $P_{M_p}$ and $P_a$, a planet mass and semi-major axis are drawn from the distribution using, 

\begin{eqnarray*}
 P_{M_p} & = & \frac{1}{N_{\rm region}} \int \limits_{\ln{M_{p_{\rm min}}}}^{\ln{M_{p}}} \int \limits_{\ln{a_{\rm min}}}^{\ln{a_{\rm max}}} C M_p^{\alpha}  \: a^{\beta} d \ln{M_p} \; d \ln{a} \nonumber \\
 & = & \frac{1}{N_{\rm region}}\frac{C}{\alpha\beta}(M_p^{\alpha} - M_{p_{\rm min}}^{\alpha})(a_{\rm max}^{\beta} - a_{\rm min}^{\beta}) \nonumber \\
 \end{eqnarray*}
 \begin{equation}
 M_p = \left(\frac{\alpha\beta}{C}\frac{P_{M_p}N_{\rm region}}{(a_{\rm max}^{\beta} - a_{\rm min}^{\beta})} + M_{p_{\rm min}}^{\alpha} \right)^{1/\alpha}
 \label{eq:massdraw}     
 \end{equation}
 
and
\begin{eqnarray*}
 P_{a} & = & \frac{1}{N_{\rm region}} \int \limits_{\ln{M_{p_{\rm min}}}}^{\ln{M_{p_{\rm max}}}} \int \limits_{\ln{a_{\rm min}}}^{\ln{a}} C M_p^{\alpha}  \: a^{\beta} d \ln{M_p} \; d \ln{a} \nonumber \\
 & = & \frac{1}{N_{\rm region}}\frac{C}{\alpha\beta}(M_{p_{\rm max}}^{\alpha} - M_{p_{\rm min}}^{\alpha})(a^{\beta} - a_{\rm min}^{\beta}) \nonumber \\
\end{eqnarray*}

\begin{equation}
a = \left(\frac{\alpha\beta}{C}\frac{P_{a} N_{\rm region}}{(M_{p_{\rm max}}^{\alpha} - M_{p_{\rm min}}^{\alpha})} + a_{\rm min}^{\beta} \right)^{1/\beta}.   
\label{eq:adraw}    
\end{equation}

This procedure is repeated by drawing new $P_a$ and $P_{M_p}$ values for each planet in the system.

\subsection{Dynamic Stability Tests} \label{sssec:stability}
A key lesson learned from the \emph{Kepler} mission is that multi-planet systems are common and often exist in compact orbital configurations \citep{Fabrycky2014,Lissauer2012}. As can be seen from Fig.~\ref{fig:PlanetsPerStar} in the SAG13 ``nominal" and ``optimistic" cases, the number of planets per star may be surprisingly large when extrapolating to large semi-major axes. This is due in part to the fact that \emph{Kepler's} primary mission observing baseline is limited (4 years) and places only a lower-limit on the total number of planets. 

Before assigning orbital elements to each planet and calculating composite stellar RV variations, we first perform an orbital stability check based on mass and semi-major axis. We analyze the long-term stability of individual members of each planetary system using mutual Hill radii, $\Delta$, by considering the circular, restricted three-body problem in a similar method as \citet{Ballard2016} and \citet{Sandford2019}. The mutual Hill radius between a pair of adjacent planets may be defined as:
\begin{equation}
\Delta = 2\left(\frac{a_{\rm out}-a_{\rm in}}{a_{\rm out}+a_{\rm in}}\right)\left(\frac{3M_{\rm Star}}{M_{p_{\rm out}}+M_{p_{\rm in}}}\right)^{1/3},   
\label{eq:delta}
\end{equation}
where $a_{\rm out}$ and $a_{\rm in}$ are the outer and inner planet semi-major axes, $M_{p_{\rm out}}$ and $M_{p_{\rm in}}$ their respective masses, and $M_{\rm Star}$ the host star mass which we set equal to 1.0 $M_{\rm \odot}$ \citep{Gladman1993,Kane2016}. \citet{Smith2009} find that an adjacent planet pair with $\Delta > 9$ can be stable for Gyr time-scales which we adopt as the minimum separation stability criteria. 

\citet{Smith2009} also  find that if a third planet on either side of the neighboring pair exists with $\Delta < 9$, such that total Hill separation of the planet triplet is $\Delta > 18$ (and provided the average Hill separation between neighboring planets remains $\Delta > 9$), then all three planets are stable. For simplicity, however, we only consider planet pairs stable if $\Delta > 9$. 

Additionally mean motion resonances may allow for planets to exist in closer, yet still stable, orbital configurations than is predicted by the stability criteria we employ \citep[e.g.,]{Lissauer2012,Gillon2017,Lissauer2011}. Consequentially our stable occurrence rates may be lower than would be found if such resonant chains were allowed.

Leading formation models such as core-accretion conclude that Jovian planets are likely to form well before Terrans \citep[]{Pollack1996,Hillenbrand2005}. To remain consistent with this concept, and given that larger planets tend to dynamically eject smaller ones in N-body simulations, we initially assume that the most massive planet in each system is stable. We then test each subsequent planet injected into the system for stability relative to the first planet using Equation \ref{eq:delta}. Any unstable planets are flagged and/or removed. This process is repeated, by checking newly synthesized planets against more massive planets to produce stable planetary systems. 

The outer-edge of the semi-major axis distribution remains poorly constrained from observations, particularly for sub-Jovian planets \citep{Suzuki2016,Crepp2011,Nielsen2019}. We find that if a sharp cutoff in the semi-major axis distribution is enforced, then an unrealistic number of planets survive the stability check at large semi-major axes. This ``edge effect'' is caused by the resulting absence of planets located exterior to the cut-off. To mitigate this effect, we temporarily use planets up to the semi-major axis cutoff of 35 au in the stability assessment but discard planets from 30-35 au in further analysis.

SAG13 is largely based on results from the \emph{Kepler} mission \citep{SAG13}. Due to the fact that completeness for close-in planets is relatively high, we implement a ``trusted'' region within which occurrence rates are matched with stably drawn planets. The outermost boundary of this ``trusted'' region is set to $a \leq 0.5$ au so as not to impact the occurrence rates at 1 au. If planets within this ``trusted'' region fail the above-mentioned stability criteria, the entire system is drawn again from the beginning until all planets located in the ``trusted'' region are dynamically stable (see also $\S$~\ref{sec:trusted}). This approach ensures that occurrence rates within 0.5 au match results from \emph{Kepler} data, since nature has already dynamically ejected the unstable planets from which the 
\emph{Kepler} demographics are empirically derived. The procedure is followed and repeated using each of the demographics sets listed in Table~\ref{tab:demosets}. 

\section{Results}\label{sec:results}

We are interested in regions in the planet mass -- semi-major axis plane where a high-contrast imaging mission could directly detect exoplanets and acquire spectra of their atmospheres. Power-law distributions are (necessarily) extrapolated to orbital distances beyond \emph{Kepler's} range and the time baselines of historic RV studies. Mutual Hill radii are calculated using the methods described above. The impact of the SAG13 demographics set (and variations thereof) on the dynamic stability of planetary systems is assessed. 

We find that dynamic instabilities govern the final number of Neptune-mass and lower-mass planets (``Terrans"). We quantify the distances from the star in which instabilities generally occur for ``optimistic," ``nominal," and ``pessimistic" scenarios. In light of these results, we develop a method for placing meaningful limits on the possible number density of planets by using ``maximally packed" orbital configurations. The results may be used with forthcoming studies to help constrain the plausible range and uncertainty of planet-yield estimates for proposed missions such as WFIRST, LUVOIR, and HabEx. 

\subsection{Dynamic Instabilities}
Our main finding is that planets drawn from SAG13 power-laws become unstable at moderate to large semi-major axes explored herein. To illustrate this, we first explore the removal of unstable planets without replacing them, as outlined in $\S$~\ref{sssec:stability}. Following this analysis, we then quantify the effects of replacing unstable planets with stable orbital configurations. We begin by separately considering three planet mass types: Jovians, Neptunes, and Terrans, as the effect of stability criteria varies between planet types.

\subsubsection{Jovian Planets $(0.225 \leq M_p / M_{\rm Jup} \leq 15)$}

\begin{figure*}
    \centering
    \makebox[\textwidth]{%
    \includegraphics[width=1.0\textwidth]{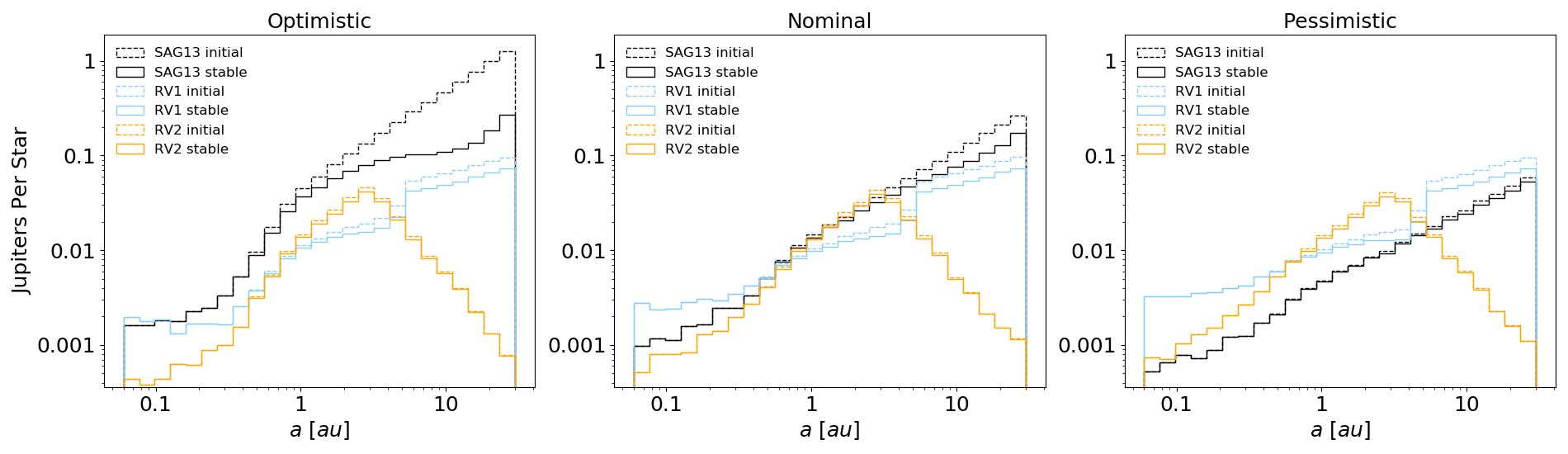}}
    \caption{Semi-major axis distribution comparing the number of Jovian-mass planets per star before (``initial") and after (``stable") dynamic stability criteria have been imposed.}
    \label{fig:1DDrawnStableJup}
\end{figure*}

The impact of the power-law distribution on the number of Jovian planets per star is shown in Fig.\ref{fig:1DDrawnStableJup}. Solid lines correspond to the ``original" population before requiring dynamic stability. Dotted lines correspond to the ``stable" population after self-consistently requiring dynamic stability for each star as described in \ref{sssec:stability}. Results are shown on a logarithmic scale to elucidate reductions in the number of Jovians per star. 

The choice of ``optimistic," ``nominal," and ``pessimistic" results in as much as an order of magnitude difference for the SAG13 distribution, with the number of Jovians per star differing most significantly (between optimistic and pessimistic scenarios) at large semi-major axes ($>1$ au). This result is not unexpected from an unbroken power-law (in semi-major axis): as the semi-major axis increases, the number of massive planets increases beyond the limit in which they can dynamically coexist. 

In the case of RV1 and RV2, the choice of ``optimistic," ``nominal," and ``pessimistic" does not directly effect the Jovian planets, it changes the distribution of the Neptune and Terran planets which are drawn from SAG13. However, there may be some differences even in the RV1 or RV2 Jovian population between the ``optimistic," ``nominal," and ``pessimistic" due to random chance or due to secondary effects imposed by the implementation of a ``trusted'' region which we discuss further in $\S$\ref{sec:trusted}.

Imposing Hill-stability criteria by iteratively injecting planets and assessing their resulting orbital architectures based on mass and spacing regulates the number of planets per star. The SAG13 distribution (black lines in Fig.\ref{fig:1DDrawnStableJup}) is suppressed much more strongly than the RV1 (blue lines) and RV2 (orange lines) distributions. As an example, the “optimistic” SAG13 distribution results in 71\% of the predicted Jupiters to be unstable. Table~\ref{tab:JupDrawnStablePercent} quantifies the percentage of predicted planets that are stable for each of the various models. 

Both the RV1 and RV2 models include power-law breaks that effectively reduce the number density of Jovian planets at larger orbital radii. As a result of a lower frequency and strength of dynamic interactions, the relative suppression of the RV1 and RV2 distributions is more modest. For RV1 and RV2, respectively ~79\% and ~93\% of Jupiters over the parameter space are found to be dynamically stable.

\begin{deluxetable}{l | c c c c }
    \tablecaption{Percentage of Jupiters predicted by each power-law set which are found to be dynamically stable }
    \label{tab:JupDrawnStablePercent}
    \tablehead{\colhead{Demographics Set} & \multicolumn{4}{c}{Semi-major axis [au]}\\
    \colhead{}&\colhead{0.06-30} & \colhead{0.06-1} & \colhead{1-10} &\colhead{10-30} }
    \startdata
        SAG13 Optimistic & 27.81 & 88.61 & 42.72 & 19.68 \\
        SAG13 Nominal & 70.08 & 97.15 & 80.11 & 63.36 \\
        SAG13 Pessimistic & 92.48 & 99.56 & 95.09 & 90.18 \\
        RV1 Optimistic & 78.36 & 96.95 & 79.18 & 75.84 \\
        RV1 Nominal & 79.24 & 97.79 & 79.73 & 76.39 \\
        RV1 Pessimistic & 79.38 & 98.07 & 79.30 & 76.50 \\
        RV2 Optimistic & 92.41 & 96.04 & 91.71 & 97.46 \\
        RV2 Nominal & 92.73 & 96.53 & 91.84 & 97.74 \\
        RV2 Pessimistic & 93.08 & 97.18 & 92.05 & 97.29 \\
    \enddata
\end{deluxetable}

Since we assume that the largest planet in each system is stable and retain the larger planet in any unstable pair, the distribution of Jovians has an impact on the stability of all other planet types as we explore in $\S$\ref{sec:NeptunesStable} and $\S$\ref{sec:EarthsStable}.

\subsubsection{Neptune Planets $(2.04 \leq M_p / M_{\oplus} \leq 75.51)$} \label{sec:NeptunesStable}

Interestingly, the influence on the semi-major axis distribution of the imposing stability criteria  becomes even more pronounced for lower mass planets. Cold Neptunes, which are concentrated near the SAG13 radius critical point of R=$3.4R_{Earth}$, have an unrealistically high occurrence rate. These cold Neptunes are unstable. Of all planet types, Cold Neptunes are being removed in the greatest number (although not in the largest percentage). 

\begin{figure*}
    \centering
    \makebox[\textwidth]{%
    \includegraphics[width=1.0\textwidth]{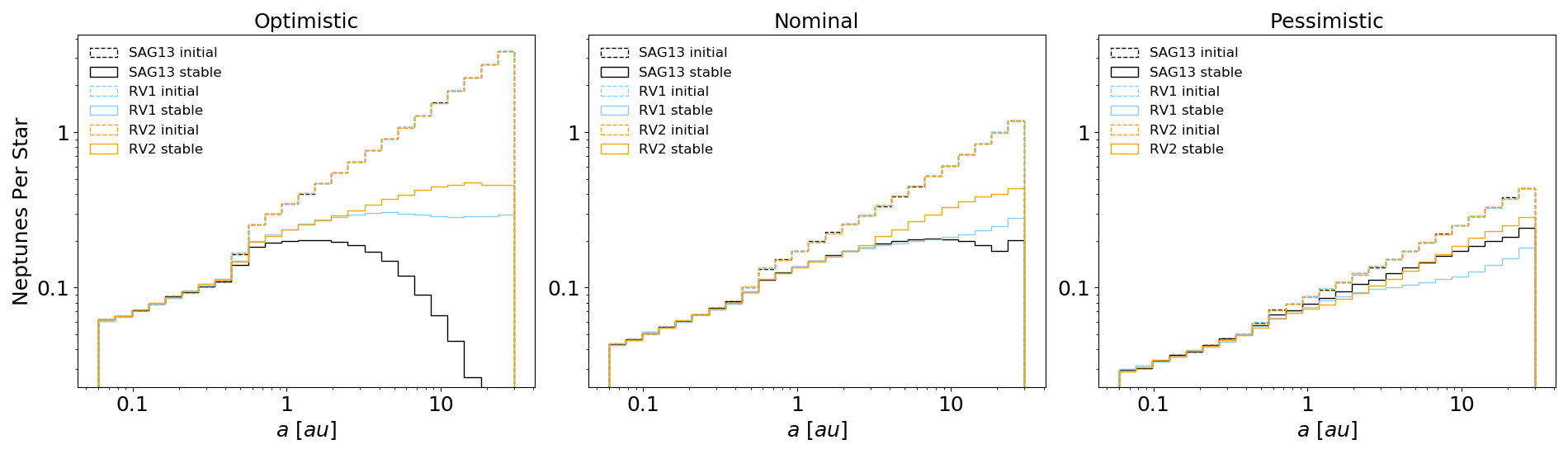}}
    \caption{Same as Fig~\ref{fig:1DDrawnStableJup}, but for Neptune-mass planets. Semi-major axis distribution comparing the number of Neptune-mass planets per star before (initial) and after dynamic stability criteria has been imposed (stable).  The initial planet populations for SAG13, RV1 and RV2 are nearly identical, because the initial demographics for Neptune-mass planets are all drawn from the same SAG13 demographics.}
    \label{fig:1DDrawnStableNep}
\end{figure*}

All three demographics sets we implement: SAG13, RV1, and RV2; use SAG13 power-law indices for Neptune and Terran mass planets. Each Neptune population begins the same, differing only by random chance, before stability criteria are applied. In Figure~\ref{fig:1DDrawnStableNep}, these initial populations are the overlapping solid lines. Interestingly however, after stability is imposed, the stable populations of Neptunes (shown in the dotted lines) differs based on the demographics set chosen. Since the only difference between the demographics sets is in the treatment of Jovians, we can see that the distribution of stable Neptunes is dependent on the initial population chosen for Jovians. We find that higher Jovian populations, such as SAG13 optimistic, allow fewer Neptunes to survive stability checks than populations with lower Jovian populations such as RV2, as might be expected.

The initial population of Jovians is not the only driving force shaping the distribution of stable Neptunes; the choice of optimistic, nominal, or pessimistic also has a major impact. Crucially here, optimistic, nominal, or pessimistic directly impacts the initial population of Neptunes (visible in the differences between the solid lines between panels of Figure~\ref{fig:1DDrawnStableNep}) as well as the initial population of Jovians for SAG13 only. In every case, pessimistic populations have a larger percentage of their Neptunes survive stability checks than nominal and optimistic populations. 

Because the choice of optimistic, nominal, or pessimistic also determines the Jovian distribution in the SAG13 only case, we can see compounding effects from both the choice of Jovian distribution (SAG13, RV1, or RV2) and the choice of optimistic, nominal, or pessimistic. For RV1 and RV2, where the choice of optimistic, nominal, or pessimistic does not effect the Jovian distribution, the occurrence rates of stable Neptunes is relatively independent of the initial choice of optimistic, nominal, or pessimistic for the initial Neptune populations. That is in those two cases, though they differ slightly from each other (blue and orange solid lines in Figure~\ref{fig:1DDrawnStableNep}), they stay fairly constant between each panel of Figure~\ref{fig:1DDrawnStableNep}. Pessimistic populations differ only slightly from this independence.

\begin{deluxetable}{l | c c c c }
    \tablecaption{Percentage of Neptunes predicted by each power-law set which are found to be dynamically stable }
    \label{tab:NepDrawnStablePercent}
    \tablehead{\colhead{Demographics Set} & \multicolumn{4}{c}{Semi-major axis [au]}\\
    \colhead{}&\colhead{0.06-30} & \colhead{0.06-1} & \colhead{1-10} &\colhead{10-30} }
    
    \startdata
        SAG13 Optimistic & 14.69 & 83.58 & 20.73 & 1.21 \\
        SAG13 Nominal & 41.96 & 92.93 & 53.76 & 21.10 \\
        SAG13 Pessimistic & 73.08 & 96.89 & 79.22 & 59.03 \\
        RV1 Optimistic & 26.77 & 87.57 & 36.69 & 11.79 \\
        RV1 Nominal & 44.45 & 93.02 & 52.98 & 26.76 \\
        RV1 Pessimistic & 59.71 & 95.60 & 64.25 & 42.77 \\
        RV2 Optimistic & 32.91 & 87.19 & 42.82 & 18.82 \\
        RV2 Nominal & 56.19 & 92.81 & 62.50 & 42.91 \\
        RV2 Pessimistic & 75.66 & 95.36 & 75.83 & 68.58 \\
    \enddata
\end{deluxetable}

\subsubsection{Terrans $(0.08 \leq M_p / M_{\oplus} \leq 2.04)$} \label{sec:EarthsStable}

\begin{figure*}
    \centering
    \makebox[\textwidth]{%
    \includegraphics[width=1.0\textwidth]{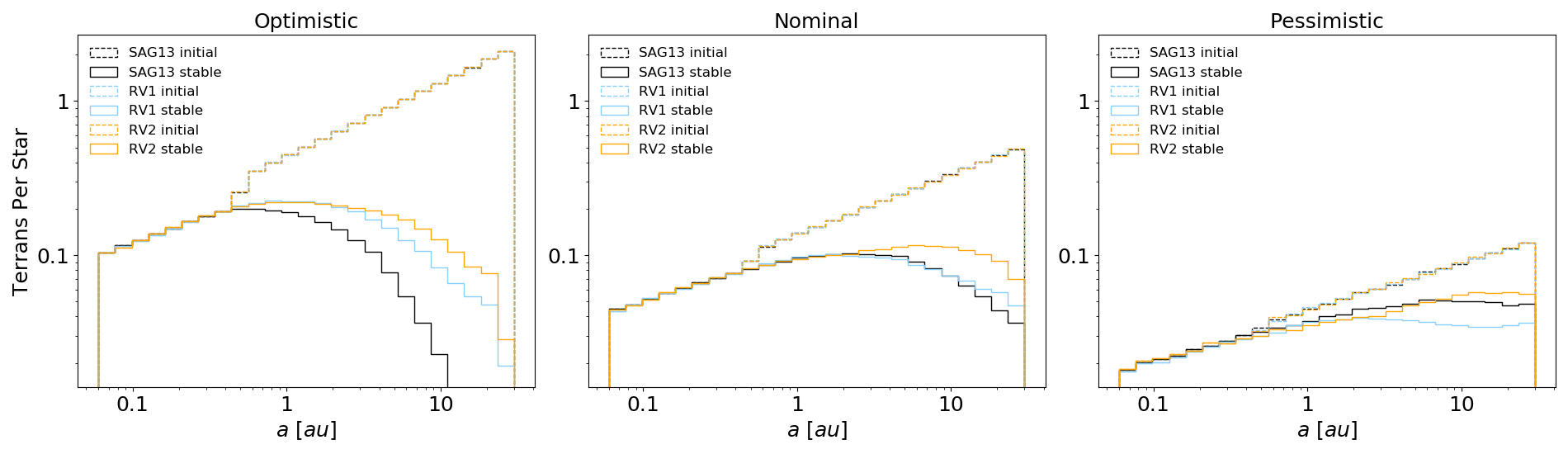}}
    \caption{Same as Fig~\ref{fig:1DDrawnStableJup}, but for Terran-mass planets. Semi-major axis distribution comparing the number of Terran-mass planets per star before (initial) and after dynamic stability criteria has been imposed (stable).  The initial planet populations for SAG13, RV1 and RV2 are nearly identical, because the initial demographics for Terran-mass planets are all drawn from the same SAG13 demographics.}
    \label{fig:1DDrawnStableEar}
\end{figure*}

Terrans have a much lower occurrence rate than Neptunes in the initial populations. Despite this, we find that the majority of Terrans predicted by SAG13 are unstable when extrapolating to large semi-major axes. Figure \ref{fig:1DDrawnStableEar} shows the semi-major axis distribution of the initial and stable populations of Terrans. When examining the Terrans only, the choice of initial Jovian distribution has little impact. We see that in all cases: optimistic, nominal, and pessimistic; the stable occurrence rates of Terrans is much lower than the initial population. Almost all Terrans are found to be unstable and removed in the furthest semi-major axis bins particularly in the optimistic case. These unstable Terrans are largely in dynamic conflicts with the large number of cold Neptunes rather than with other small mass planets. 
\begin{deluxetable}{l | c c c c }
    \tablecaption{Percentage of Terrans predicted by each power-law set which are found to be dynamically stable }
    \label{tab:EarDrawnStablePercent}
    \tablehead{\colhead{Demographics Set} & \multicolumn{4}{c}{Semi-major axis [au]}\\
    \colhead{}&\colhead{0.06-30} & \colhead{0.06-1} & \colhead{1-10} &\colhead{10-30} }
    
    \startdata
        SAG13 Optimistic & 16.55 & 78.85 & 13.89 & 0.44 \\
        SAG13 Nominal & 39.58 & 90.05 & 43.21 & 12.31 \\
        SAG13 Pessimistic & 68.37 & 95.34 & 71.30 & 46.34 \\
        RV1 Optimistic & 21.24 & 81.34 & 21.52 & 2.85 \\
        RV1 Nominal & 39.96 & 90.12 & 42.45 & 14.20 \\
        RV1 Pessimistic & 57.95 & 94.24 & 58.53 & 33.12 \\
        RV2 Optimistic & 22.97 & 81.26 & 23.90 & 4.48 \\
        RV2 Nominal & 45.85 & 89.83 & 48.48 & 22.66 \\
        RV2 Pessimistic & 68.80 & 93.85 & 67.69 & 53.41 \\
    \enddata
\end{deluxetable}

\subsubsection{Entire Population}

\begin{figure*}
    \centering
    \makebox[\textwidth]{%
    \includegraphics[width=1.0\textwidth]{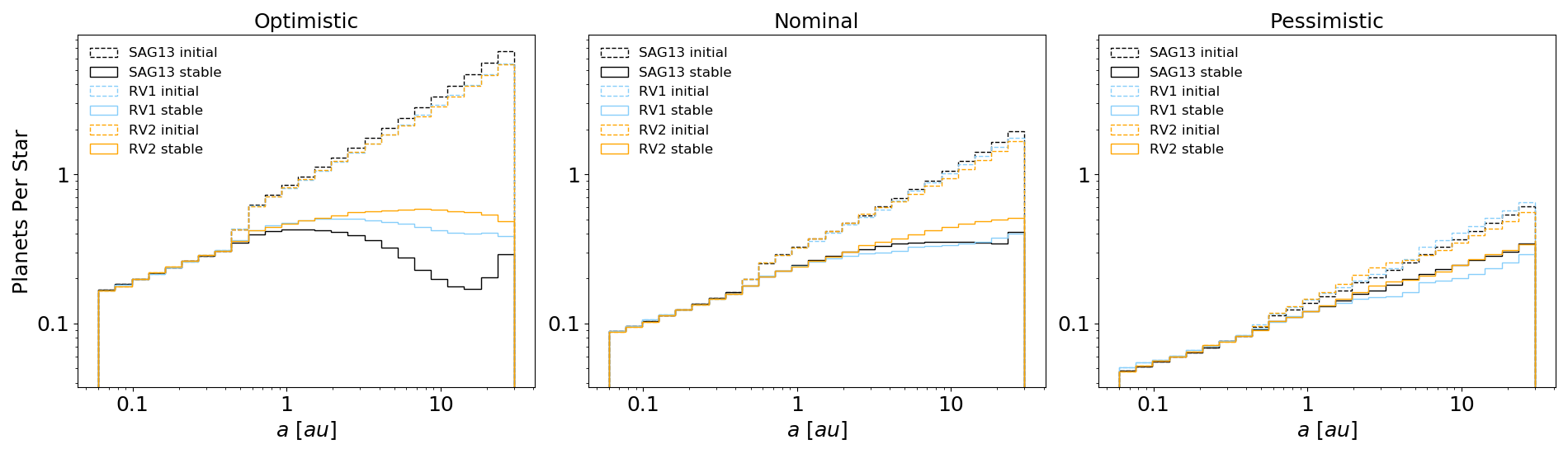}}
    \caption{Semi-major axis distribution of initially predicted  and stablity checked with no redraws populations for optimistic, nominal, and pessimistic sets. Within a 0.5 au "trusted region", we match all three distributions by redrawing the entire population for a star if unstable.}
    \label{fig:1DDrawnStable}
\end{figure*}

We next examine the effect of stability on the combined population of all planets. Figure~\ref{fig:1DDrawnStable} shows the semi-major axis distribution of the entire stable population in comparison to the initial (non-stable) population.  The number of planets per star for stable systems is substantially lower than the initially drawn planets outside of the ``trusted'' region, where they are forced to match \emph{Kepler} catalogs. The difference is particularly pronounced for cold planets when extrapolating to large semi-major axes and for optimistic occurrence rates. Fundamentally this indicates that a large number of initially generated cold planets are unstable. Thus naively extending SAG13 occurrence rates results in unrealistic systems which have far more planets than dynamic stability would allow for.  

\begin{figure*}
    \centering
    \makebox[\textwidth]{%
    \includegraphics[width=1.0\textwidth]{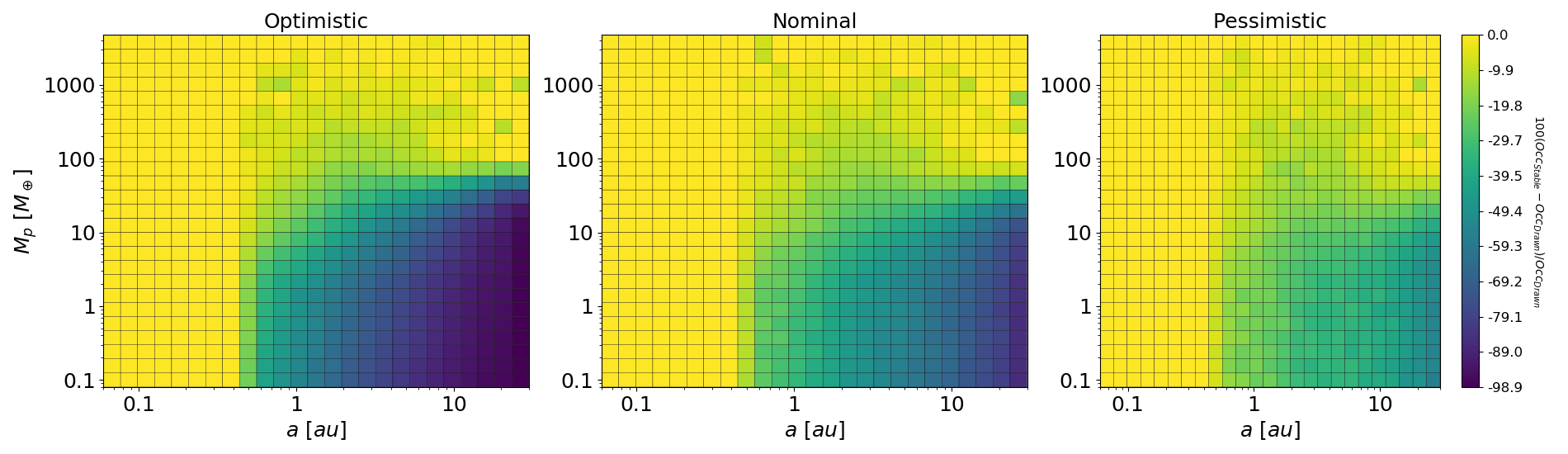}}
    \caption{Percent difference between initially predicted and stablity checked with no redraws populations (as a percentage of initially predicted) in two-dimensional mass and semi-major axis space for the RV2 set.}
    \label{fig:PerDiffDrawnStable}
\end{figure*}

In Figure~\ref{fig:PerDiffDrawnStable}, we show the percent difference between the initial and stable populations. As previously discussed, the greatest percent differences occur in cold Terrans with almost all such planets being lost to stability checks in the optimistic case. Due to lower numbers of initial planets in the pessimistic case, a lower percentage of cold planets are removed by stability checks in the pessimistic case although still with $\sim40$\% reduction. 

\begin{figure*}
\centering
    \makebox[\textwidth]{%
    \includegraphics[width=1.0\textwidth]{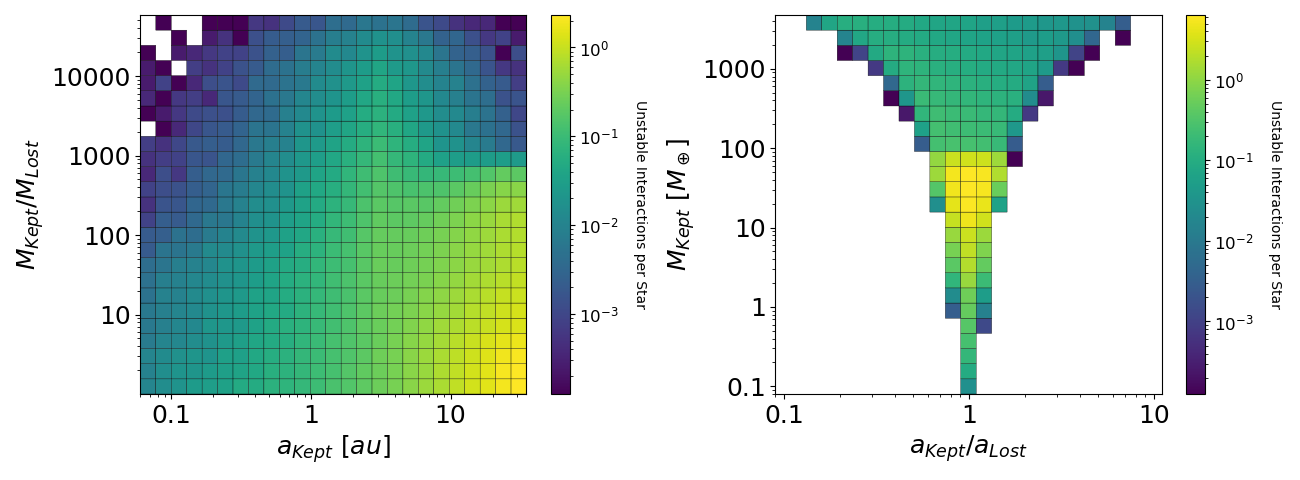}}
    \caption{Comparison of demographics of dynamically conflicting individual planet pairs from the initial stability check with the color bar scaling as interactions per star (with 100,000 stars) for the RV2 Nominal set. These plots include unstable interactions from systems which are later redrawn entirely due to the instability of a ``trusted'' planet, as described in $\S$~\ref{sssec:stability}. Also shown are the interactions involving planets between 30-35 au which are not included in further analysis due to an ``edge effect'' also described in $\S$~\ref{sssec:stability}. On left is shown the mass ratio of the kept planet to the lost planet as a function of semi-major axis of the kept planet. Most lost planets are removed by a planet slightly more massive than itself and at large semi-major axis. The sub-figure at right shows the ratio of semi-major axis between the kept and lost planet as a function of the mass of the kept planet. Most planets are removed by Neptune-mass planets and in close proximity in semi-major axis space; however larger mass planets remove planets that differ more in distance.}
\label{fig:KeptPlanets}
\end{figure*}

It is also worth examining which planets have the largest effect on limiting the occurrence rates. In figure \ref{fig:KeptPlanets}, we look at the relationship between mass and semi-major axis of planets in dynamic conflicts, specifically in this case for the RV2 Nominal set. On the left panel, we see that most of the ``kept'' planets in such conflicts are in the largest semi-major axis bins and that conflicts are mainly between planets that have a mass ratio less than 10. In the right panel, we see that most of the ``kept'' planets have a mass in the Neptune range and that the ratio of ``kept'' to ``lost'' planet semi-major axes is close to 1 except in the case of the most massive ``kept'' planets. Fundamentally this has several implications:

\begin{enumerate}
    \item Cold Neptunes have the largest impact on stability due to their large occurrence rate
    \item Planets are mostly removed by slightly more massive and mostly nearby planets
    \item Large planets (Jovians) have lower impact on stability than Neptunes because Jovians are less common but Jovians impact a wider range of semi-major axes. 
\end{enumerate}

While in any sufficiently large random draw it is likely at least a few planets would be drawn in unstable configurations by chance, the large amount of planets removed from the ``initial'' populations by the first stability check shows we are stability-limited at large semi-major axes. Non-random trends in both the planets that are removed, and the larger planets that are kept, impact planet occurrence rates of their systems. We could assume that this stability-regulated planet population that we have generated represents a nominal approximation of the true unknown planet population at large semi-major axes. However, simply removing unstable planets from the initially drawn planet population is logically flawed to a certain extent. While cold exoplanets occurrence rates are somewhat unconstrained, rates closer-in are more reliable and based on mature, real systems which are already known to be stable; subtracting unstable planets drawn in these regions without any attempted replacement then artificially lowers the occurrence rate in known regions with respect to to the true demographics by a small percentage. In other words, we "double subtract" dynamically unstable planets -- first by nature, and second from the stability check. This must also be true at large semi-major axes, and thus the planet population at large semi-major axes may be under-estimated by an unknown amount from the actual (unknown) demographics.  

We next investigate what bounds we can infer on planet populations at large semi-major axes when we are stability-limited. The lower bound for the demographics of Neptune and Terran planets at large semi-major axes is not empirically constrained, and technically could be zero. We can use the concept of dynamical stability, however, to place a meaningful upper bound for exoplanet occurrence rates of Neptune and Terran planets at large semi-major axes.


\subsection{A Maximally Packed Planet Population}
\label{sssec:maxpack}

We generate demographics that result from maximally packing planetary systems. As a comparison to both the initial and stability-checked demographics, we have created an additional maximally packed scenario where each system is dynamically packed with the maximum number of planets via redraws from the original power-law distributions. This scenario represents an upper-bound on cold planet occurrence rates. It is likely that nature does not always produce maximally packed systems; as an example, \citet[]{Mordasini2018} predicts an the occurrence rate of planets at small semi-major axes that is higher than Kepler statistics \citep{Mulders2019,Fernandes2019}. However, in this case, the orbital dynamics in \citet[]{Mordasini2018} are simulated for a less than the age of the Kepler systems. Additionally, as previously mentioned, resonant chains of planets can yield planetary systems that are dynamically stable and more compact than the mutual Hill radii criterion employed herein \citep[e.g.,]{Lissauer2012,Gillon2017,Lissauer2011}

\begin{figure}
    \centering
    \includegraphics[width=0.5\textwidth]{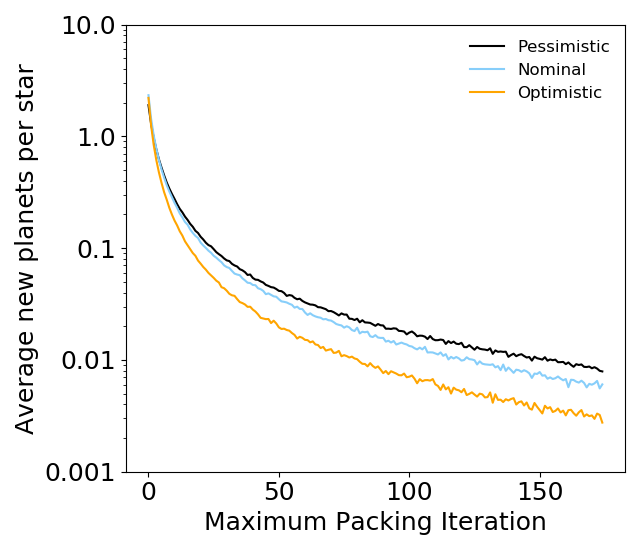}
    \caption{Average planets per star added with each iteration of the maximum packing procedure for the optimistic, nominal, and pessimistic populations. Each population reaches $<0.01$ planets per star within 175 iterations.}
    \label{fig:PlanetsPerLoop}
\end{figure}

In order to add additional planets to a system, we recalculate an expected number of planets for each power-law region by integrating over the parameter space outside of the trusted region (beyond 0.5 au) via eq. \ref{eq:Npl}. Again we calculate an integer number of new planets via eq. \ref{eq:poisson}. For each new planet, we draw a mass and semi-major axis via eqs. \ref{eq:massdraw} and \ref{eq:adraw} in the same manner as the original planets were drawn. If these new planets are found to be dynamically stable with: 1) every original stable planet, and 2) with any planets previously added by this maximally packing procedure, then the new planets are retained and added to a population we denote as maximally dynamically packed.

For each star, we redraw new planets and attempt to add them to the system 175 times, each time adding in any stable planets. This number of iterations was chosen because we find at this point that on average $<0.01$ planets per star are added in a single loop indicating that $<1$\% of stars were able to fit an additional planet in the last iterations. 

A secondary, although minor, effect on this maximum packing procedure stems from the type of power-law used. The number of planets attempted in each iteration varies based on the expected number of planets. Thus more optimistic demographics sets, which have a higher expected number of planets and so have more planets attempted in each iteration, reach a lower percentage of stars able to add a planet in the final iterations than more pessimistic demographics sets. This can be clearly seen in Figure~\ref{fig:PlanetsPerLoop}, where the optimistic case (orange) reaches a lower average of new planets per star than the pessimistic case (black).  

\subsection{Impact of Maximum Packing}

\begin{figure*}
    \centering
    \makebox[\textwidth]{%
    \includegraphics[width=1.0\textwidth]{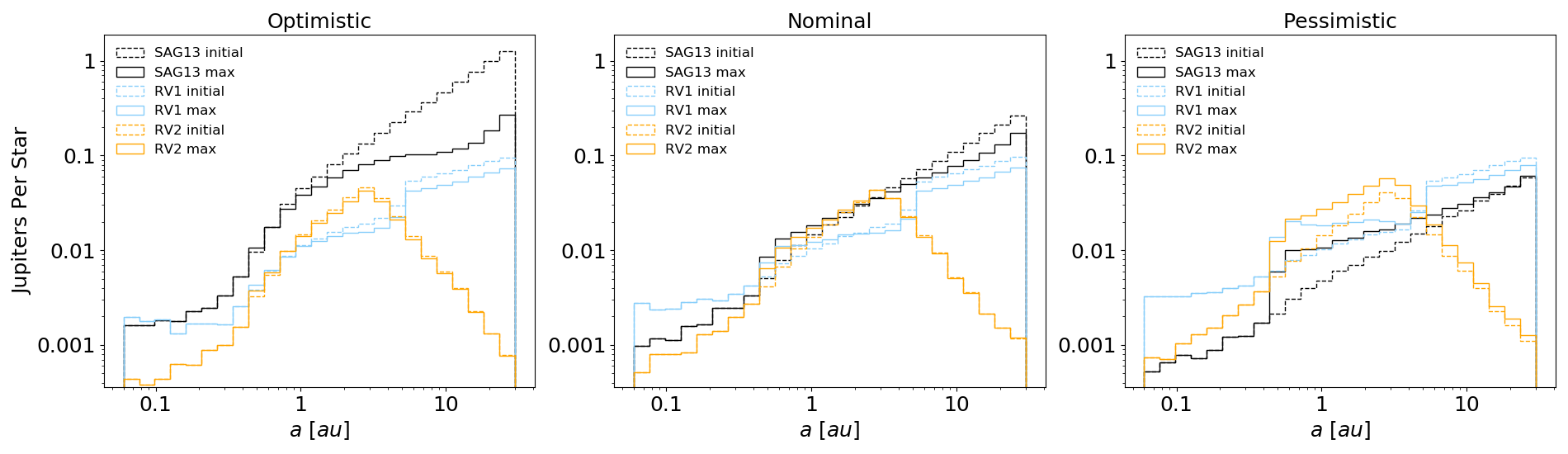}}
    \caption{Semi-major axis distribution of initially predicted and maximally dynamically packed populations of Jovians.}
    \label{fig:1DDrawnMaxJup}
\end{figure*}

\begin{figure*}
    \centering
    \makebox[\textwidth]{%
    \includegraphics[width=1.0\textwidth]{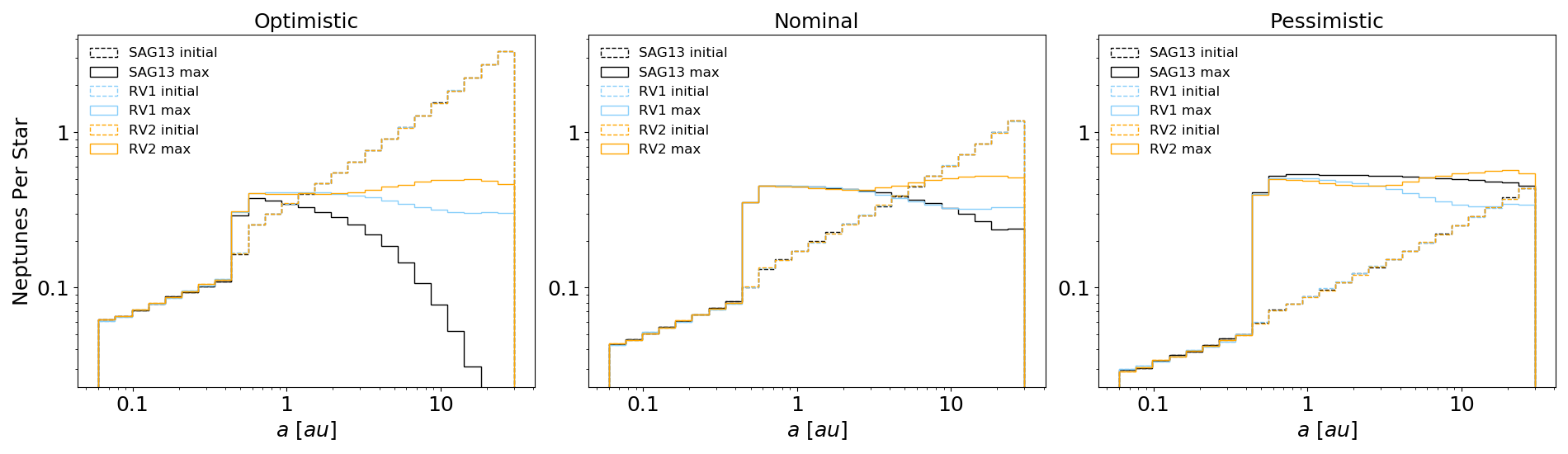}}
    \caption{Semi-major axis distribution of initially predicted and maximally dynamically packed populations of Neptunes.}
    \label{fig:1DDrawnMaxNep}
\end{figure*}

\begin{figure*}
    \centering
    \makebox[\textwidth]{%
    \includegraphics[width=1.0\textwidth]{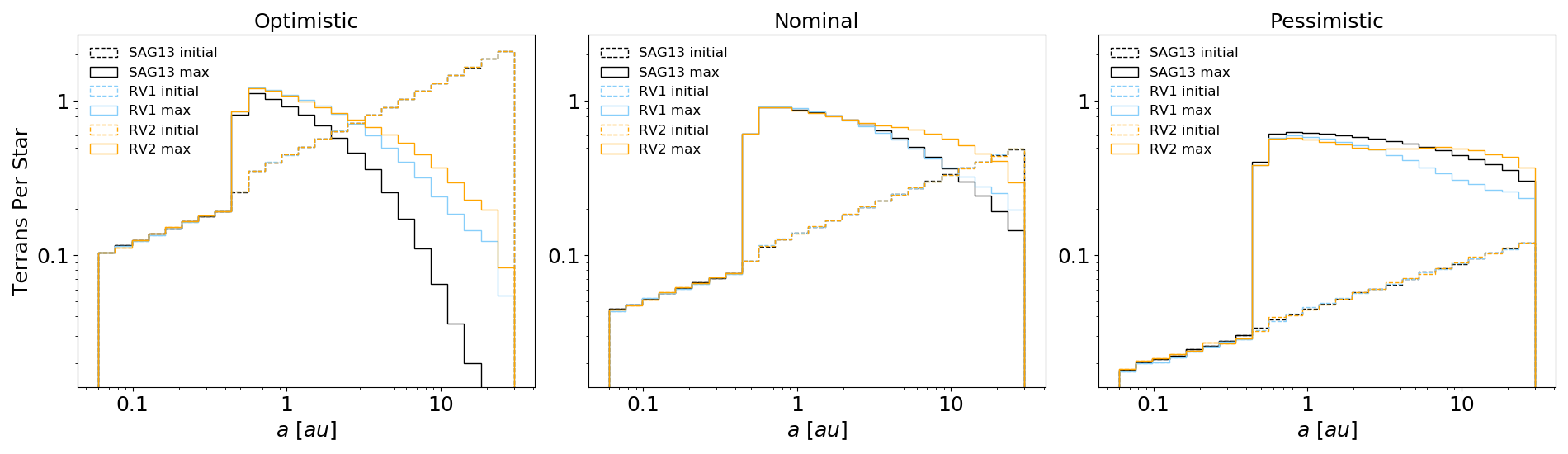}}
    \caption{Semi-major axis distribution of initially predicted and maximally dynamically packed populations of Terrans.}
    \label{fig:1DDrawnMaxEar}
\end{figure*}

\begin{figure*}
    \centering
    \makebox[\textwidth]{%
    \includegraphics[width=1.0\textwidth]{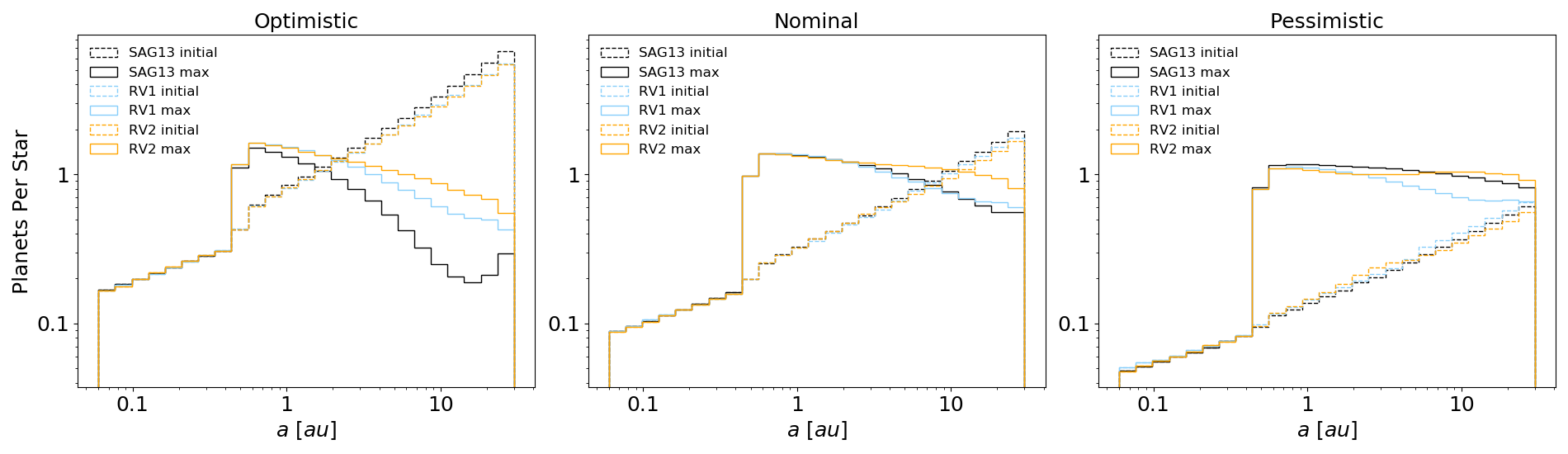}}
    \caption{Semi-major axis distribution of initially predicted and maximally dynamically packed populations optimistic, nominal, and pessimistic sets. Within a 0.5 au "trusted region", we match all three distributions by redrawing the entire population for a star if unstable.Beyond the point where the non-stability checked distribution exceeds the maximally packed distribution expected occurrence rates cannot be safely extended without a correction for stability.}
    \label{fig:1DDrawnMax}
\end{figure*}

To assess the impact of maximum packing, we begin again by dividing each population into planet types: Jovian, Neptunes, and Terrans. In Figure~\ref{fig:1DDrawnMaxJup} we see that we are not able to maximally pack the Jovians to fit the initially predicted occurrence rates. In fact, very few Jupiters are added in any set over the ``stable with no redraws'' values. 

For Neptunes, the impact of maximum packing is entirely different. In Figure~\ref{fig:1DDrawnMaxNep} we see that maximum packing cannot match initial cold occurrence rates in the optimistic and nominal cases. However, maximum packing exceeds initially predicted values for close in rates and over the entire semi-major axis space for many pessimistic cases. 

A similar situation is found for Terrans. In Figure~\ref{fig:1DDrawnMaxEar}, we see maximum packing greatly exceeds initially predicted values close-in but still falls short (in the optimistic and nominal cases) for cold planets.

Looking at the entire population of planets, in Figure~\ref{fig:1DDrawnMax}, we see that the maximally packed systems have overall lower cold planet occurrence rates than predicted by the initially drawn systems for the nominal and optimistic cases. The ``cross-over'' point where initially drawn exceeds maximally packed varies between the nominal and optimistic cases, occurring at $\sim10$ au and $\sim2$ au, respectively. However, for the SAG13 pessimistic case, the maximally packed scenario exceeds initial over the entire parameter space. 

\begin{figure*}
    \centering
    \makebox[\textwidth]{%
    \includegraphics[width=1.0\textwidth]{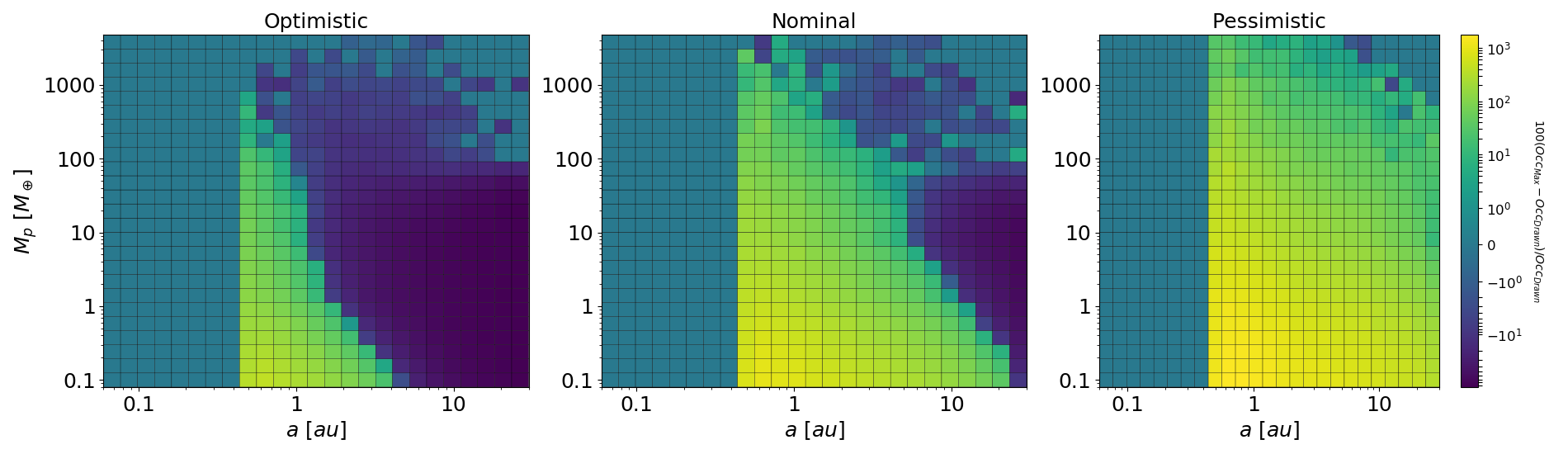}}
    \caption{Percent difference between initially predicted and maximally packed populations (as a percentage of initially predicted) in two-dimensional mass and semi-major axis space for the RV2 set.}
    \label{fig:PerDiffDrawnMax}
\end{figure*}

A trend, exists not only in semi-major axis space but in two-dimensional mass-semi-major axis space. Figure~\ref{fig:PerDiffDrawnMax} shows that the majority of additional planets in the maximally packed case are sub-Earth-sized and near the boundary of the trusted region. Even maximally packed systems cannot reach the levels of Cold Neptunes predicted by SAG13 optimistic and nominal rates. The rates of Cold Terrans as well in maximally packed populations, in the optimistic case, are found to be far below the levels predicted by SAG13. In the pessimistic case, we see that maximum packed populations greatly exceed SAG13 over nearly the entire parameter space. 

\subsection{Notes on the ``Trusted'' Region}\label{sec:trusted}

One major drawback of implementing a ``trusted'' region is that within the ``trusted'' region the initial drawn population may not accurately reflect the power-law. We require that any planets within the ``trusted'' region be stable, even in the initial population, and randomly redraw the entire system if they are not. Through this requirement, we slightly bias the results within that region to lower occurrence rates than would be predicted from the power-law. Ideally this the trusted region would have very low percent differences between the power-law and the initial population, varying only due to random fluctuations. This however, is not what we see, particularly in the SAG13 optimistic case. In this case, we see that ``initial'' population rates are lower than predicted by the power-law in the trusted region particularly for more massive planets.

\begin{figure*}
    \centering
    \makebox[\textwidth]{%
    \includegraphics[width=1.0\textwidth]{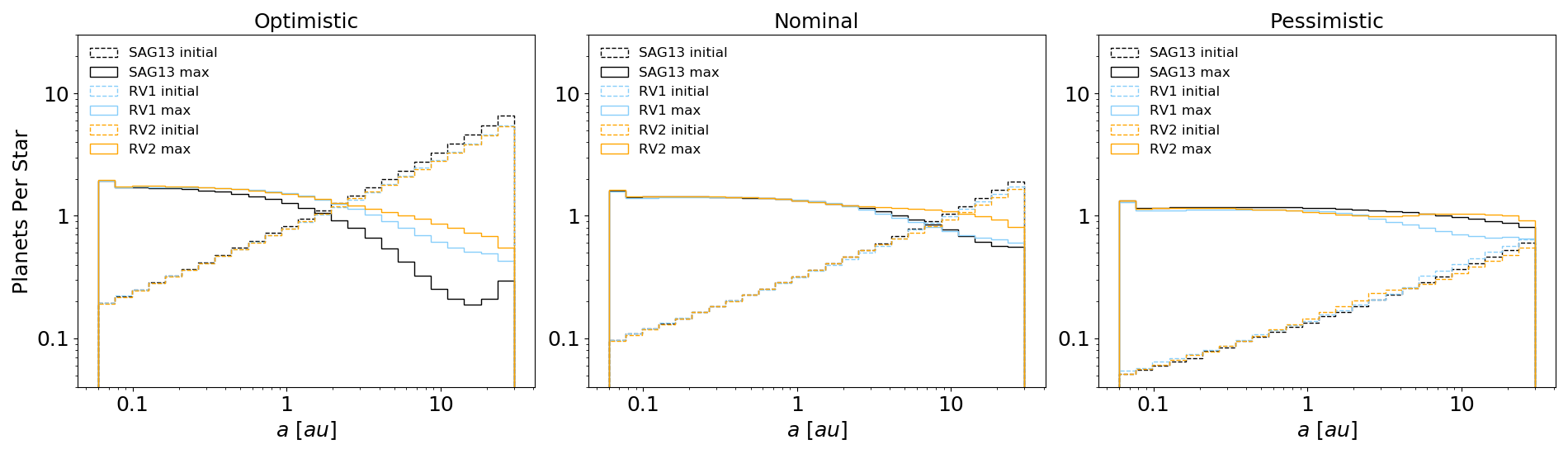}}
    \caption{Semi-major axis distribution of initially predicted and maximally dynamically packed populations with no ``trusted'' region.}
    \label{fig:1DDrawnMaxNoTrusted}
\end{figure*}

Additionally, there is a large jump in maximally packed systems at the edge of the ``trusted'' region as seen in Figure~\ref{fig:1DDrawnMax} at 0.5 au. A logical question then follows as to what the occurrence rate would be if such a ``trusted'' region were removed. Allowing planets to be added down to the minimum mass however produces a large amount of close-in planets, as shown in Figure~\ref{fig:1DDrawnMaxNoTrusted}. These planets could have reasonably been expected to be detected by {\it Kepler} if the planets reflect true occurrence rates. 

Because {\it Kepler}-based SAG13 has lower predictions than maximally packed for close-in planets, we know that SAG13 does not predict maximally packed systems. This limits maximum packing as a realistic model of planetary occurrence rates.

Lastly, in Table \ref{tab:kopparapu}, we present a comparison of the various datasets within the planetary definition framework outlined in Kopparapu et al. 2018. Many of the trends we have previously discussed are visible in this table. 

\begin{deluxetable*}{@{\extracolsep{10pt}}l c c c c c c c c c}
    \tablecaption{Number of Planets per star for characteristic Radius-Insolation bins based on \citet[][]{Kopparapu2018}. Notably, Stable with no redraws shows a sharp decrease from the predicted population in cold rocky planets for both the optimistic and nominal cases. Also of note, after applying maximum packing, the optimistic case does not always predict more planets of every type than the nominal case or even the pessimistic case (see, for example, the Cold sub-Neptunes bin). This is likely due to higher initial values after the initial stability check and preferential sorting to keep the previously stable planets in any unstable pair.}
    \label{tab:kopparapu}
    \tablehead{\colhead{Planet type} & \multicolumn{3}{c}{RV2 Optimistic} & \multicolumn{3}{c}{RV2 Nominal} & \multicolumn{3}{c}{RV2 Pessimistic}\\
    \cline{2-4} \cline{5-7} \cline{8-10}
    \colhead{}&\colhead{Initial}&\colhead{No Redraws}&\colhead{Max}&\colhead{Initial}&\colhead{No Redraws}&\colhead{Max}&\colhead{Initial}&\colhead{No Redraws}&\colhead{Max}}
    
    \startdata
        Hot Rocky& 1.81 & 1.46 & 3.98 & 0.63 & 0.57 & 2.59 & 0.22 & 0.21 & 1.46 \\
        Warm Rocky& 1.06 & 0.45 & 2.28 & 0.31 & 0.19 & 1.83 & 0.09 & 0.07 & 1.13 \\
        Cold Rocky& 7.47 & 1.14 & 4.29 & 1.88 & 0.75 & 4.88 & 0.5 & 0.32 & 3.55 \\
        Hot super-Earths& 0.87 & 0.74 & 1.22 & 0.43 & 0.39 & 1.05 & 0.20 & 0.19 & 0.86 \\
        Warm super-Earths& 0.57 & 0.29 & 0.65 & 0.22 & 0.15 & 0.70 & 0.09 & 0.07 & 0.66 \\
        Cold super-Earths& 4.21 & 0.81 & 1.37 & 1.44 & 0.61 & 1.86 & 0.51 & 0.33 & 1.98 \\
        Hot sub-Neptunes& 0.65 & 0.58 & 0.82 & 0.44 & 0.42 & 0.85 & 0.28 & 0.27 & 0.90 \\
        Warm sub-Neptunes& 0.44 & 0.28 & 0.44 & 0.24 & 0.18 & 0.53 & 0.13 & 0.1 & 0.63 \\
        Cold sub-Neptunes& 3.54 & 0.92 & 1.16 & 1.63 & 0.81 & 1.60 & 0.77 & 0.54 & 2.01 \\
        Hot sub-Jovians& 0.09 & 0.08 & 0.11 & 0.06 & 0.06 & 0.12 & 0.04 & 0.04 & 0.13 \\
        Warm sub-Jovians& 0.13 & 0.10 & 0.14 & 0.07 & 0.06 & 0.14 & 0.04 & 0.03 & 0.16 \\
        Cold sub-Jovians& 3.03 & 1.05 & 1.13 & 1.29 & 0.78 & 1.15 & 0.57 & 0.44 & 1.16 \\
        Hot Jovians& 0.06 & 0.05 & 0.06 & 0.04 & 0.04 & 0.06 & 0.03 & 0.03 & 0.06 \\
        Warm Jovians& 0.10 & 0.09 & 0.10 & 0.05 & 0.05 & 0.08 & 0.03 & 0.03 & 0.08 \\
        Cold Jovians& 2.11 & 1.32 & 1.34 & 0.77 & 0.62 & 0.73 & 0.31 & 0.27 & 0.51 \\
    \enddata
\end{deluxetable*}

\begin{figure*}
    \centering
    \bf{Optimistic}
    \gridline{\fig{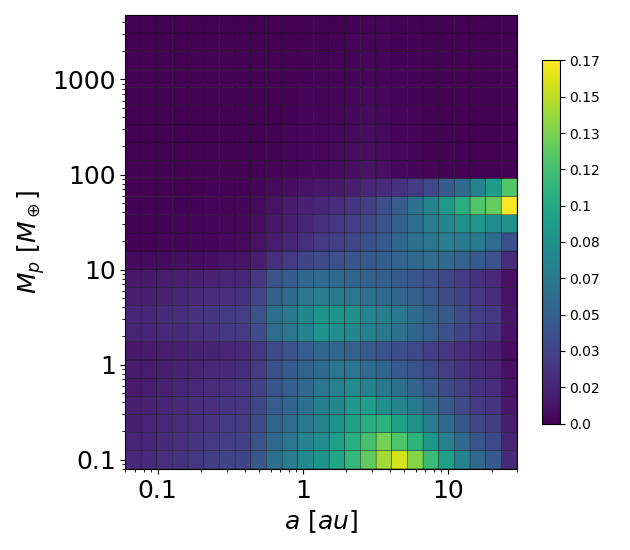}{0.40\textwidth}{}
            \fig{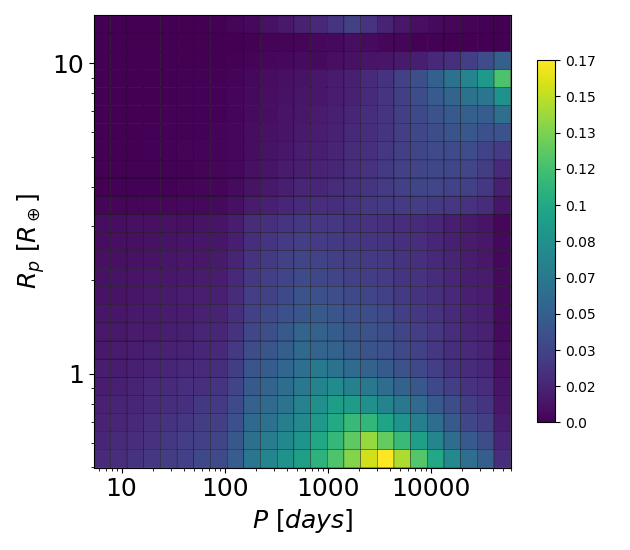}{0.40\textwidth}{}}
    \bf{Nominal}
    \gridline{\fig{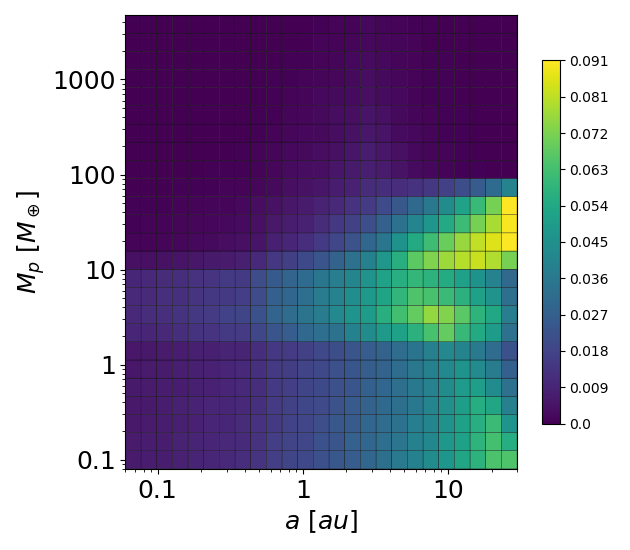}{0.40\textwidth}{}
            \fig{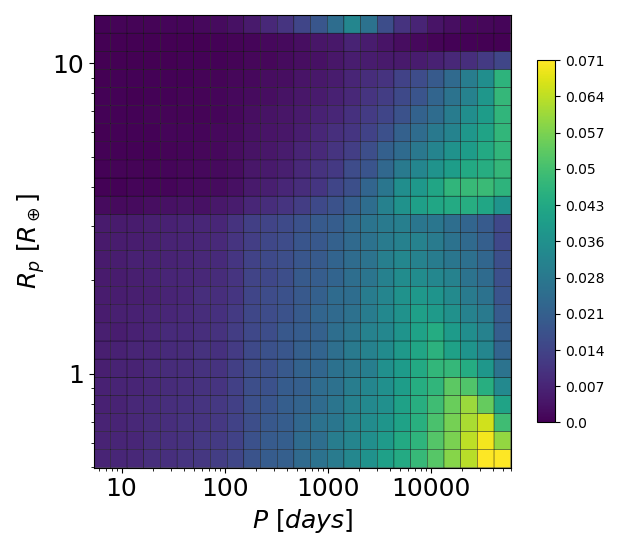}{0.40\textwidth}{}}
    \bf{Pessimistic}
    \gridline{\fig{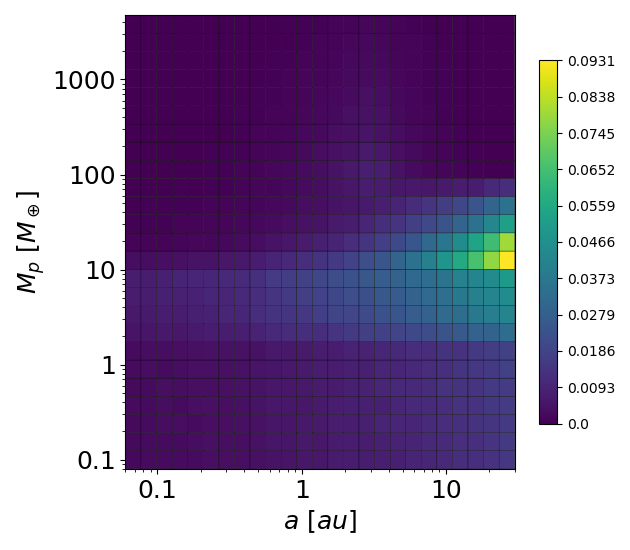}{0.40\textwidth}{}
            \fig{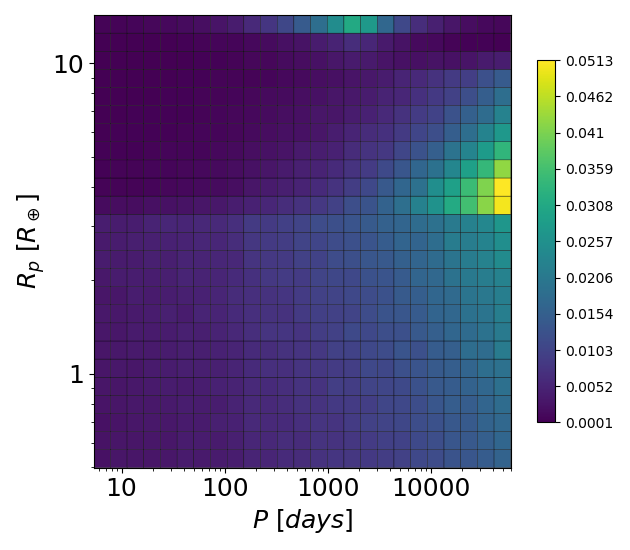}{0.40\textwidth}{}}
    \caption{Resulting occurrence rates, derived from the RV2 optimistic, nominal and pessimistic sets, after taking the cell-by-cell minimum of initial and maximally packed populations. For the lower 3 figures, this cell-by-cell minimum is done after planetary parameters are translated back to the original parameter space of SAG13 (radius, period). Also see note in on this translation in the Appendix. These occurrence rates could be utilized as an update to SAG13 informed by dynamic stability.}
    \label{fig:FinalOcc}
\end{figure*}

\section{Discussion and Concluding Remarks}
\label{sec:conclusions}

Our understanding of the make-up and orbital architecture of the inner-most portions of other solar systems is largely informed by observations made by NASA's \emph{Kepler} space mission. Not only are terrestrial planets common, but multi-planet systems are prevalent \citep[]{Fabrycky2014,Lissauer2012}. Such extensive statistical information was unavailable only several years ago. 

What the {\it Kepler} results imply for future direct imaging missions however is less clear and warrants further study through modeling and numerical simulations of space-craft capabilities. The time baseline and sensitivity of {\it Kepler} offered reasonably complete period distributions for planets comparable to and larger than the Earth with orbits located interior to the habitable zones of solar-type stars \citep{Kopparapu2018}. Analyzing trends in planet occurrence as a function of semi-major axis, and allowing for small extrapolations thereof, suggests that many more planets could reside at orbital distances of one to several astronomical units. Nevertheless, the value of $\eta_{\oplus}$ remains uncertain. This ``warm" planet population, which does not experience extreme stellar radiation levels,\footnote{The eccentricity distribution could further impact habitability \citep{Williams2002}.} may not be thoroughly studied until WFIRST commences its microlensing survey \citep{Penny2019}.  

RV technologies will continue to improve in the interim \citep{NAS2018}. If single measurement precisions approach $\sigma=10$ cm/s, then the most promising high-contrast imaging targets could be pre-selected based on the presence of Doppler accelerations \citep{Crepp2012,Crepp2014}. Even with current technologies, WFIRST itself ---assuming it is equipped with a coronagraph--- would benefit from the measurements of precursor RV surveys, particularly when considering the ever-more accurate determination of Keplerian orbital elements and ephemeris refinement \citep{Kane2013,Kane2018}. Ground-based RV observations would enable full orbital solutions and dynamical mass estimates \citep{Crepp2016,Crepp2018}.

The same methods currently employed from the ground using adaptive optics could be implemented with terrestrial planets from space. Proposed missions such as HabEx, LUVOIR, and WFIRST Starshade Rendezvous could offer several orders of magnitude deeper contrast than WFIRST CGI, making direct detection of entire planetary systems routine \citep{Lacy2019}. Because exposure time of these future flagships will be a valuable commodity, information regarding each star’s planetary system that allows us to better optimize observations could significantly improve efficiency, and potentially data quality. Any boosts in scientific output have yet to be quantified; they are also quite sensitive to the value of $\eta_\oplus$.

The first steps towards quantifying the benefits of using RV measurements for future imaging missions involves compiling occurrence rate information. Developing models, sooner rather than later, that attempt to ``stitch together" planet occurrence rates using disparate techniques will allow much-needed time for analysis of systematic offsets, selection effects, observational biases and self-consistency checks. Inevitable extrapolations into previously uncharacterized regions of parameter space are necessary to best-design the cadence and many other aspects of precision RV surveys, which in turn are needed years before any imaging system is launched \citep{HowardFulton2016}.

In order to gain a handle on the number density of planets, and corresponding yield estimates plausible directly imageable planets, we have analyzed a list of ``demographics sets" compiled from literature references and NASA ExoPAG SAG13. We directly compare three demographics sets which all utilize SAG13 rates for Terran and Neptune planets but which differ in the occurrence rate distribution of Jovians. ``Stitching together" these power-law-based demographics and converting to common units necessarily requires both extrapolation and interpolation to incorporate the entire parameter space of interest.

The plausible range of masses, orbits, and overall number of terrestrial planets is highly uncertain in regions just exterior to those searched by \emph{Kepler}. As a constraint on the various mass and semi-major axis demographics models, we performed a first-order analysis of synthesized multi-planetary systems by assessing dynamic stability using mutual Hill radii. Even a marginal extrapolation of the planet distributions to orbital radii of interest to high-contrast imaging missions shows that secular gravitational interactions would disrupt system stability on timescales of Gyr or shorter. 

In particular, we find that an over-density in the number of Neptune-mass planets causes strong and frequent interactions that affect the orbits of yet lower-mass planets, should they exist. Somewhat ironically, we find that the various Jovian distributions explored have only a marginal effect on the stability of lower mass planets. Since in all cases Neptune-mass planets are far more common then Jovian-mass planets, the Neptunes have a much larger impact on the stability of low-mass planets than the Jovians. Thus the choice of ``optimistic", ``nominal", or ``pessimistic" rates for Neptune and lower-mass planets has the most pronounced impact on system stability.  

To compile realistic systems for mission yield calculations, we have attempted to ``squeeze the problem" of unknown, cold planet occurrence rates by placing meaningful upper limits and lower limits on the value; this involved developing a procedure for self-consistently generating configurations that consist of ``maximally packed" orbits. The populations of planets which are assessed for dynamical stability but where unstable planets are not replaced represent a somewhat ``nominal'' scenario. Conversely the maximally dynamically packed systems represent an upper limit on planet occurrence. We find that at $a\sim10$ au for the ``nominal" case (or ~2 au for the optimistic case), extensions of SAG13 exceed the occurrence rates of maximally packed systems, particularly for small planets. Beyond this point, stability should be accounted for in yield calculations of cold Terran planets.  

A hybrid model that uses SAG13 for close-in planets, yet replaces cold planets (where SAG13 over-predicts occurrence rates) with a maximally packed population, might be used as a reasonable model for planet occurrence. In principle, this approach could be achieved by taking the cell-by-cell minimums of the radius-period histograms of the initial population and the maximally packed population (Figure~\ref{fig:FinalOcc}). In the pessimistic case, this is simply the initial population since the initial population never has higher occurrence per bin than the maximally packed population. Notably, we find that \emph{Kepler} statistics do not predict maximally packed systems at small semi-major axes. Thus an assumption of maximum packing systems at large semi-major axes may still over-predict the true number of cold planets. As a caveat to this analysis, we have not allowed for resonant ``chains of planets," which could further increase occurrence rates \citep{Gillon2017,Lissauer2011}. 

As we discussed in $\S$~\ref{sec:methods}, we assume that the average number of planets per star is over all stars and we do not introduce a multiplicative factor of the fraction of stars with planets. Were this done, the average number of planets per star for stars with planets would rise to meet the total overall average predicted by the occurrence rates. In doing so, the stability constraints we employ would have an even larger impact on these now additionally over-predicted systems.

In this paper, we have utilized a simplistic stability criteria that does not depend on effect of eccentricity or mutual inclinations in the stability of systems. High eccentricities, in particular, would be expected to decrease the number of planets able to be maximally packed in stable systems. In order to place meaningful upper limits on planet occurrence rates we have assumed circular orbits which would allow for higher global occurrence rates. This assumption likely holds true for smaller mass (Neptune and Terran) planets which have been found to have lower average eccentricities than Jovians \citep{Kane2012}. There is both observed and simulated evidence that high eccentricity systems likely result from planet-planet scattering which would have a larger effect in systems with Jovians \citep{Kane2014,Carrera2019}.

It is not yet known whether Jovian occurrence rates at large and small semi-major axes are coupled. In this paper, we have considered two additional Jovian distributions in comparison to the SAG13 results which we have labeled RV1 and RV2. RV1 is based on a combination \citet{Bryan2016} and \citet{Cumming2008}. As mentioned in $\S$\ref{sec:sets}, our use of the \citet{Bryan2016} demographics for cold Jupiters assumes a decoupled distribution between hot and cold populations noticeable in the occurrence rate break at the transition from \citet{Cumming2008} to \citet{Bryan2016} at 5 au. Our RV2 set, based on \citet{Fernandes2019}, does not make this assumption and instead assumes a turnover in Jovian occurrence rates. In our analysis, the comparative occurrence rates of Jovians did not have an overwhelming impact on the stable occurrence rates of Neptunes or Earths. Although generally very high Jovian occurrences, such as SAG13 optimistic, allowed for fewer stable small planets than either RV1 or RV2 sets. Ultimately our final occurrence rates, shown in Figure \ref{fig:FinalOcc}, are derived from RV2 sets.

It is worth considering the extent to which ground based detections might improve the efficiency of multi-billion dollar NASA flagship missions such as WFIRST. A precursor RV survey could potentially decrease the amount of time required for WFIRST to characterize planets, saving time in the 5-year nominal mission lifetime and boosting planet yields. A relevant previous preliminary analysis of this topic is in \citet{Dressing2019}. This Astro2020 White Paper calculated the impact that such a precursor RV survey would have on a LUVOIR-A scale direct imaging mission. \citet{Dressing2019} found, in the case of an extremely precise and complete RV survey down to 8 cm/s, that the initial search more time of a direct imaging mission would be reduced by 44\%. This initial search is predicted to 71\% or 48\% of the LUVOIR-A or HabEx mission lifetimes respectively \citep{Roberge2018,Gaudi2018}. Thus a significant reduction in the initial search would greatly boost mission scientific yields by allowing for a focus on planet characterization. A precursor RV survey would have the additional benefit of supplying mass measurements, which both reduce the amount of time in verifying false positives as well as allow for the interpretation of planetary spectra by providing constraints on a surface gravity estimate. 

The techniques we have developed for planet occurrence may be used to generate planetary systems for joint RV and direct imaging survey simulations. Forthcoming papers in the series will present results using these methods for NASA's WFIRST and LUVOIR / HabEx missions. 

This research is supported by a contract from the NASA Jet Propulsion Laboratory and Exoplanet Exploration Program in support of the HabEx Mission Study Science and Technology Definitions Team. SDD acknowledges support by the National Science Foundation Graduate Research Fellowship Program under Grant No. DGE-1841556 and the 2016 NSF REU program at Notre Dame under Grant No. PHY-1559848. Any opinions, findings, and conclusions or recommendations expressed in this material are those of the authors and do not necessarily reflect the views of the National Science Foundation. JRC acknowledges support from the NSF CAREER award No. AST-1654125.

\begin{appendix}
 In this appendix, we detail the methods we employed to ``stitch together'' demographics power laws of various formulations into a uniform mass-semimajor axis parameter space. In equation \ref{eq:intergratedgeneral}, we stated that the average number of planets per star per mass-semimajor axis region in our general notation is:
 
 \begin{equation*}
    N_{\rm region} = \frac{C_{X,Y,Z}}{\alpha_X \beta_Y (\ln{Z})^2}(X_{\rm max}^{\alpha_X} - X_{\rm min}^{\alpha_X})(Y_{\rm max}^{\beta_Y} - Y_{\rm min}^{\beta_Y})
\end{equation*}

In order to determine this value for particular mass-semimajor axis boundaries as we do in equation \ref{eq:Npl}, it is necessary to convert $X_{\rm max}$, $X_{\rm min}$, $Y_{\rm max}$, $Y_{\rm min}$, $\alpha_{X}$, and $\beta_{Y}$ to Earth mass and au units. The conversions are listed below for common conversions. Substitution of these conversions into equation \ref{eq:intergratedgeneral} for particular mass-semimajor boundaries of interest will allow for the calculation of $N$ directly from the original power-law values.

\begin{align*}
    &X_{mass,Jup} = \frac{1.2668653e17}{3.986004e14} Mp \qquad \text{(IAU definition)}\\
    &\alpha_{mass,Jup} = \alpha_{Mp} \\
    &X_{radius,Earth} = C_{\text{ChenKipping}}Mp^S \\
    &\qquad \qquad \text{where $C_{\text{ChenKipping}}$ and $S$ are the relevant power law} \\
    & \qquad  \qquad \text{constant and exoponent, from \citet{ChenKipping2017}} \\
    &\alpha_{radius,Earth} = S\alpha_{Mp} \\
    &\\
    &Y_{period,days} = 365.25a^{1.5} \\
    &\beta_{period,days} = 1.5 \beta \\
    &Y_{period,years} = a^{1.5} \\
    &\beta_{period,years} = 1.5 \beta
    &
\end{align*}

The expected value, $N_{\rm region}$, is in equation~\ref{eq:intergratedgeneral}, except in the case where $\alpha$ or $\beta$ are zero. For SAG13, we utilize Chen \& Kipping 2017 to convert planet radius to mass with the appropriate break between Terran and Neptune type planets \citep{ChenKipping2017}. Because the Chen \& Kipping Jovian region is characterized by a negative mass exponent, radii between 11.3 and 14.3 $R_{Earth}$ cannot be uniquely converted to masses. Accordingly we do not consider the Jovian mass-radius relation and we extend the Neptune mass-radius relation to the upper mass limit of $15 M_J$. This is only necessary in the case of SAG13 only demographics and not for the cases of the RV1 or RV2 sets. 

For the panels on the right-hand side of Figure \ref{fig:FinalOcc}, individual simulated planets are translated back to the original parameter space of SAG13 (radius and period) before the cell-by-cell minimum of the initial and maximally packed histograms is taken. The translation from semimajor axis to period is straightforward: $P = 365.25a^{1.5}$. The translation from mass to radius utilizes the Chen \& Kipping mass-radius relation. In this case, to avoid the degeneracy in the function from 11.3 to 14.3 $R_{Earth}$, all planets below $0.414M_{Jup}$ utilize either the Chen \& Kipping Terran or Neptune relation (as appropriate) and all planets above $0.414M_{Jup}$ (entirely drawn in this case from the Fernandes occurrence rates) utilize the Jovian relation.  

\end{appendix}

\bibliography{Demographics.bib}

\end{document}